\documentclass[nonatbib,tocpage]{cablab}

\usepackage{algorithmic}
\usepackage{algorithm}
\usepackage{textcomp}
\usepackage{verbatim}
\usepackage{tikz}
\usetikzlibrary{positioning,arrows.meta,calc,fit,backgrounds}
\usepackage[numbers]{natbib}

\newcommand{\IEEEPARstart}[2]{#1#2}
\newcommand{\IEEEpubidadjcol}{}
\makeatletter
\NewEnviron{IEEEkeywords}{\global\let\cab@keywords\BODY}
\makeatother
\titleformat{\paragraph}[runin]
  {\sffamily\bfseries\normalsize\color{cabAccent}}{}{0pt}{}

\renewcommand{\thesection}{\Roman{section}}
\renewcommand{\thesubsection}{\thesection-\Alph{subsection}}
\renewcommand{\thesubsubsection}{\thesubsection\arabic{subsubsection}}

\title{Geometric Algebra for Neural Networks:\\ A Survey}
\shorttitle{Geometric Algebra for Neural Networks: A Survey}

\author[1]{Tao Zhong}
\author[2]{Kaison Fong}
\author[1]{Keqin Wang}
\author[1]{Christine Allen-Blanchette}
\affiliation[1]{Princeton University}
\affiliation[2]{Stanford University}

\begin{abstract}
Geometric algebra (GA) unifies geometric primitives and transformations in a single algebraic framework. Neural networks designed around this framework often represent features as multivectors and constrain layer operations to respect the algebraic structure, enabling principled treatment of geometric transformations and, in many cases, built-in equivariance. This survey reviews geometric algebra neural networks, including their building blocks, equivariant properties, application domains, empirical performance, and software support. We provide an algebra-selection guide, an empirical comparison of GA-based and other equivariant architectures, and a failure-mode analysis covering training pathologies, deployment constraints, and performance limitations. Across the surveyed literature, the strongest evidence for GA-based architectures appears in settings with explicit geometric symmetries and heterogeneous geometric entities; evidence for broader advantages in parameter efficiency, interpretability, and dense prediction is comparatively limited and uneven. We therefore conclude that geometric algebra neural networks are not universally superior to well-tuned non-GA equivariant baselines, and outline open problems in algebra selection, numerical stability, scalable implementation, and reproducible evaluation.
\end{abstract}


\date{\today}
\project{https://cab-lab-princeton.github.io/AwesomeGANN}
\metadata[\faGithub\ Paper list]{\href{https://github.com/CAB-Lab-Princeton/AwesomeGANN}{\nolinkurl{github.com/CAB-Lab-Princeton/AwesomeGANN}}}
\correspondence{\email{tzhong@princeton.edu}, \email{ca15@princeton.edu}}

\begin{document}
\maketitle

\section{Introduction}
\IEEEPARstart{N}{eural} networks (NNs)~\cite{goodfellow2016deep,lecun2015deep} have achieved broad success across perception and representation learning~\cite{krizhevsky2012imagenet,he2016deep}, but they often treat geometric data through scalar or tensor parameterizations that obscure its underlying structure. This limitation is especially acute in modern perception and reasoning systems, where models must jointly manipulate heterogeneous geometric entities, including points, pixels, poses, rays, lines, planes, rigid motions, and cross-sensor transformations. In conventional pipelines, these entities are typically encoded using disparate representations and connected through ad hoc conversions. When such information is flattened into unstructured feature vectors, relationships such as incidence, orientation, metric structure, and transformation equivariance are no longer explicit~\cite{wu2020comprehensive}. As a result, models may require extensive data augmentation to approximate basic symmetries~\cite{cohen2016group,shorten2019survey}, generalize poorly to unseen geometric configurations~\cite{zhang2016understanding,esteves2018learning}, and provide limited insight into the geometric computations they perform~\cite{bau2017network}. These difficulties point less to a lack of model capacity than to a mismatch between the structure of geometric data and the representations through which many neural architectures process it.

Clifford algebra~\cite{dorst2009geometric,doran2003geometric,clifford1871preliminary,hestenes2015space}, often called geometric algebra (GA) in applied settings, offers a compelling response to this representational fragmentation. It provides an algebraically coherent framework in which diverse geometric entities, including points, lines, planes, circles, and spheres, and transformations, including rotations, translations, reflections, and dilations, can be represented within a common language. This contrasts with conventional pipelines in which directions, rotations, rigid motions, and projective transformations are often handled by separate formalisms such as vector algebra, quaternions, homogeneous coordinates, or transformation matrices. By encoding geometric structure directly into representations and operations, GA-native models can impose geometric priors that are difficult to express in unstructured scalar feature spaces. In selected settings, these priors have been associated with improved data efficiency~\cite{brehmer2023geometric,ruhe2023geometric} and more interpretable intermediate representations~\cite{pepefengbo}. For the pattern analysis and machine intelligence community, GA is especially relevant when a task requires joint reasoning over multiple geometric entity types, as in 3D vision, embodied perception, multi-sensor fusion, pose estimation, and manipulation. In these contexts, a unified algebraic framework can simplify layer design and reduce representational fragmentation.

Research on GA-based neural networks has grown rapidly alongside geometric deep learning (GDL), which seeks to encode geometric priors and symmetries directly into model architectures~\cite{bronstein2021geometric}. Recent GA-native models have appeared across biology~\cite{wagner2024generating,liu2025clifford}, physics~\cite{chen2024machine}, and computer science~\cite{ruhe2023geometric,ruhe2023clifford}, reflecting the appeal of GA as a common language for structured geometric computation. At the same time, the literature remains difficult to synthesize because methods differ in their choice of algebra, notation, equivariance target, architectural primitives, and evaluation protocol. This survey addresses that fragmentation by organizing GA-native neural networks according to shared mathematical foundations, design patterns, empirical evidence, and open limitations.

\IEEEpubidadjcol
Within GDL, many successful architectures are designed around particular data structures or symmetry groups. Graph neural networks (GNNs)~\cite{wu2020comprehensive} operate on relational and non-Euclidean data, while group-equivariant~\cite{cohen2016group,esteves2018learning} and steerable~\cite{cohen2016steerable} convolutional networks encode prescribed transformation laws for images, volumetric fields, and related grid-structured signals. These approaches provide strong inductive biases, but their mathematical formulations are often specialized to particular symmetry groups, domains, or representation types. GA-native networks offer a complementary route: they use multivector representations and Clifford-algebra operations to express geometric entities and transformations within a common algebraic language. This perspective has been instantiated across graphs~\cite{zhong2025graph,liu2022geometric}, point clouds~\cite{xu2024point,brehmer2023geometric}, manifolds~\cite{quackenbush2024geometric}, and fields~\cite{brehmer2023geometric}. However, this generality does not automatically imply empirical superiority. As Section~\ref{sec:non-ga-comparison} and Section~VI detail, mature non-GA equivariant architectures remain competitive in many settings, and a central aim of this survey is to assess where GA-native structure provides a genuine advantage.

This survey distinguishes itself from prior work through a focused treatment of what we term \emph{GA-native} neural networks: models whose learnable representations and core operations act directly on multivectors or structured grade-wise GA objects. We exclude works that use GA only for preprocessing, feature engineering, or coordinate conversion while performing learning in conventional scalar or tensor spaces. Borderline cases, such as conformal-GA input embeddings followed by otherwise standard networks, are included only when GA operations remain part of the learned forward map. In contrast to general GA surveys~\cite{hitzer2024current,hitzer2024survey,hitzer2013applications,wang2019geometric,macdonald2017survey} and broader GDL surveys~\cite{heidari2024geometric,zhang2023systematic,sanborn2024beyond,basheer2026symmetry}, we emphasize GA-native architectures and synthesize empirical conclusions only when methods are compared on shared tasks, datasets, and metrics. Fig.~\ref{fig:survey-positioning} situates the survey within this adjacent literature. Our main contributions are:
\begin{figure*}[t]
\centering
\includegraphics[width=\textwidth]{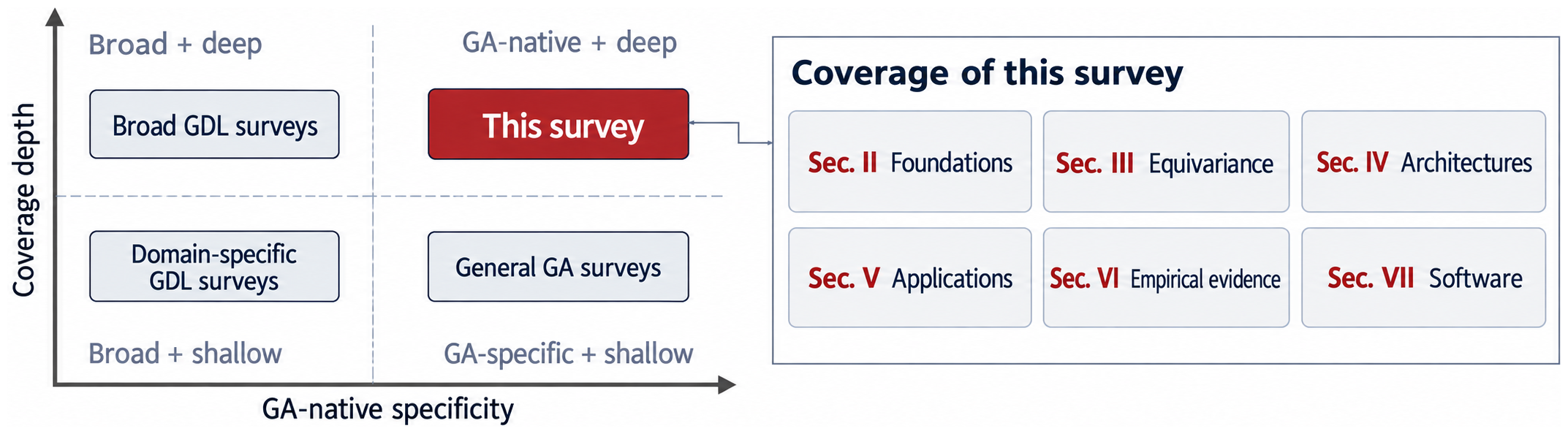}
\caption{Positioning of this survey relative to adjacent literatures. Existing surveys either treat GA broadly across applications or GDL broadly across formalisms; this survey focuses on GA-native neural networks while covering foundations, architectures, applications, empirical evidence, and software.}
\label{fig:survey-positioning}
\vspace{-8pt}
\end{figure*}
\begin{itemize}[leftmargin=*]
    \item A unified taxonomy of GA-native neural networks, organized by algebra choice, multivector representation, layer operation, equivariance target, architecture family, and application domain.
\item An algebra-selection guide comparing Euclidean, projective, conformal, and spacetime algebras in terms of their geometric primitives, transformation groups, computational costs, and modeling trade-offs.
\item A critical synthesis of empirical evidence comparing GA-native models with non-GA equivariant and geometric deep learning baselines on shared tasks, datasets, and metrics.
\item A failure-mode and implementation analysis covering numerical stability, training pathologies, scalability, software support, deployment constraints, and reproducibility.
\item A set of open problems and research directions for principled architecture design, benchmark construction, efficient implementation, and reliable evaluation.
\end{itemize}

\subsection{Survey Methodology}
\label{sec:methodology}
This survey follows a structured literature-search protocol. We searched major academic databases and preprint repositories using combinations of Clifford- and geometric-algebra terminology with neural-network and deep-learning terms, augmented by domain-specific modifiers such as vision, molecular, physics, and robotics. We also performed backward and forward citation tracking from foundational GA-NN papers~\cite{brehmer2023geometric,ruhe2023clifford,brandstetter2022clifford}. Preprints that appeared at peer-reviewed venues by the time of writing are treated as published work.

\emph{Inclusion criteria.} A work is included if its learnable forward map operates directly on multivectors or structured grade-wise GA objects, which we refer to as the \emph{GA-native} criterion. Works that use GA only for preprocessing, feature engineering, or coordinate conversion, while carrying out the main learning process in conventional scalar or tensor spaces, are excluded from the core corpus. Borderline cases, such as CGA-based input embeddings followed by otherwise standard networks, are included only when GA operations remain part of the learned computational graph.

\emph{Exclusion criteria.} We exclude pure-mathematics GA papers without a learning component, tutorial introductions, and works operating exclusively on quaternions or dual quaternions without broader Clifford-algebraic structure. Non-archival workshop abstracts that lack sufficient methodological detail for assessment are also omitted.

After deduplication and screening, the final corpus comprises approximately 80 GA-native works, together with approximately 40 contextual references spanning general GA, GDL, and non-GA baseline methods. In the empirical synthesis of Section~VI, we grade each matched benchmark setting as \emph{Strong}, \emph{Moderate}, or \emph{Preliminary} according to the reliability of the comparison. This grade reflects our confidence in the evidence rather than the magnitude or direction of any reported performance gap. It is set by the number of independent replications, the maturity and relevance of the benchmark, the availability of strong non-GA baselines, and the consistency of the evaluation protocol. A central outcome of this synthesis is partly negative: outside settings with explicit geometric symmetries or heterogeneous geometric entities, the current literature does not yet support broad claims of GA-native superiority, owing to sparse matched benchmarks, limited dense-prediction results, and heterogeneous evaluation practices.

\begin{figure*}[t]
\centering
\includegraphics[width=\textwidth]{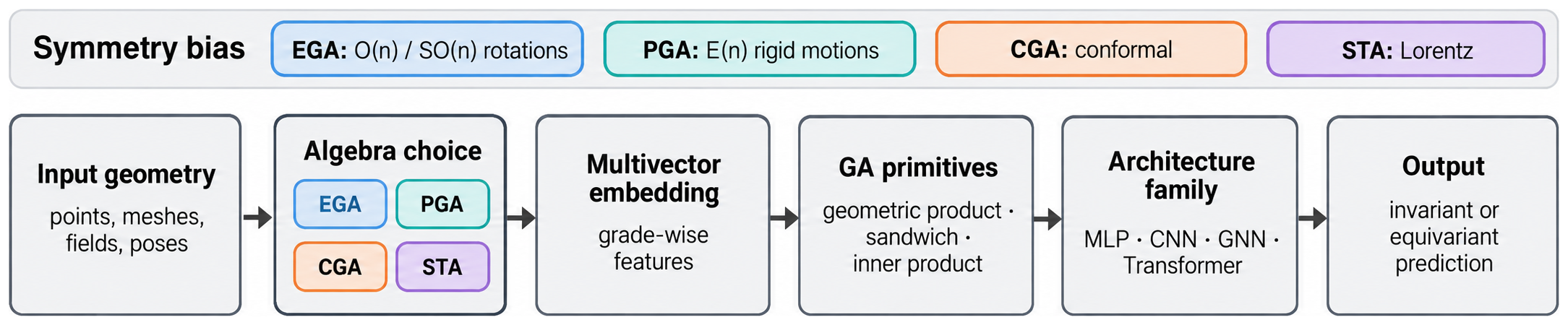}
\caption{Generic GA-native neural-network pipeline. The choice of algebra fixes the geometric vocabulary and symmetry bias; multivector embeddings are then processed by GA primitives and assembled into architecture families matched to downstream tasks.}
\label{fig:ga-pipeline}
\vspace{-8pt}
\end{figure*}

The remainder of the paper is organized as follows, with Fig.~\ref{fig:ga-pipeline} providing a compact roadmap of the design space. Section~II revisits foundational GA concepts with emphasis on their relevance to NN design and provides an algebra-selection decision aid. Section~III discusses the GDL principles realized by GA-native architectures and compares them with non-GA equivariant alternatives. Section~IV reviews GA-native layers and architecture families. Section~V surveys applications. Section~VI synthesizes empirical performance with evidence-strength grading. Section~VII covers software support and implementation practice. Finally, Section~VIII outlines open problems and concludes the survey.

\section{Foundational Concepts: GA for NNs}
\subsection{GA Fundamentals} 
\indent This section summarizes the GA concepts needed for neural-network design. Let \(V\) be an \(n\)-dimensional real vector space equipped with a quadratic form of signature \((p,q,r)\), where \(n=p+q+r\). The associated geometric algebra \(\mathbb{G}_{p,q,r}\) is the associative algebra generated by basis vectors \(\{e_i\}_{i=1}^n\) whose squares are \(+1\), \(-1\), or \(0\), depending on whether they correspond to positive, negative, or null directions of the metric. The geometric product of basis vectors satisfies \(e_i e_j=-e_j e_i\)~\cite{roelfs2023graded}. 

The fundamental element of GA is the \textit{multivector}, which is a linear combination of components of different \textit{grades}. In an \(n\)-dimensional algebra, a multivector \(M\) can be decomposed as $M = \langle M \rangle_0 + \langle M \rangle_1 + \langle M \rangle_2 + \cdots \langle M \rangle_{n},$ where $\langle M \rangle_k$ denotes the grade-$k$ component. Grade-0 elements are scalars, grade-1 elements are vectors, grade-2 elements are bivectors, grade-3 elements are trivectors, and the highest-grade element is the pseudoscalar \(e_1e_2\cdots e_n\). A \textit{$k$-blade} is an outer product of $k$ linearly independent vectors and represents an oriented \(k\)-dimensional subspace. Every multivector can be expressed as a linear combination of basis blades. Since there is one basis blade for each subset of the \(n\) basis vectors, \(\mathbb{G}_{p,q,r}\) has $\sum_{k = 0}^{n} \binom{n}{k} = 2^{n}$ basis blades.

The geometric product is the central operation of GA. For vectors \(x,y\in V\), it decomposes into symmetric and antisymmetric parts \(xy = x\cdot y + x\wedge y\). The inner product (first term) measures metric similarity, while the outer product (second term) constructs oriented subspaces. In GA-native neural networks geometric-product layers can mix grades and introduce new grades unless the architecture constrains them.

A multivector $R$ is called a \textit{versor} if it can be written as a geometric product of invertible vectors $v_1 v_2 \cdots v_r.$ Versors generate geometric transformations, vectors, through sandwich actions of the form
\(x \mapsto R x R^{-1}\). Even versors include rotors and motors, which are widely used to represent rotations and rigid motions in GA-native architectures. Thus, multivectors provide the representation space, the geometric product provides the primitive algebraic operation, and versors provide a compact mechanism for encoding transformation structure.

\subsection{Key Algebras}
\indent GA-native neural networks differ not only in architecture but also in the choice of underlying algebra. The most common choices in the current literature are Euclidean GA (EGA), projective GA (PGA), conformal GA (CGA), and spacetime algebra (STA). This choice determines the available basis blades, native geometric objects, transformation groups, equivariance constraints, and computational cost (see Table~\ref{tab:ga_compare}). Selecting the appropriate algebra therefore remains an active design question~\cite{de2024euclidean,pepe2024staresnet,vasconcelos2024correspondence}: EGA is compact but lacks native translations, PGA represents rigid motions naturally, CGA adds conformal and distance-aware structure at higher dimension, and STA is tailored to Lorentzian geometry.

\begin{table*}[t]
\caption{Principal algebras used in GA-native neural networks. Canonical references are listed in the rightmost column.}
\label{tab:ga_compare}
\centering
\footnotesize
\setlength{\tabcolsep}{4pt}
\renewcommand{\arraystretch}{1.15}
\begin{tabularx}{\textwidth}{@{}lll>{\hsize=.9\hsize\raggedright\arraybackslash}X>{\hsize=1.1\hsize\raggedright\arraybackslash}X>{\raggedright\arraybackslash}p{3.3cm}@{}}
\toprule
Algebra & Signature & Native symmetry & Native objects / actions & NN design takeaway & Representative works \\
\midrule
EGA & $\mathbb{G}_{3,0,0}$ & $O(3)$/$SO(3)$ & vectors, bivectors, rotors; rotations, reflections & compact and stable; no native translations, often needs centering & CGENN~\cite{ruhe2023clifford}, GCAN~\cite{ruhe2023geometric} \\
PGA & $\mathbb{G}_{3,0,1}$ & $E(3)$ & planes, lines, points, motors; rigid motions & native isometries for meshes and poses; inner product not distance-aware & GATr~\cite{brehmer2023geometric}, GAGrasp~\cite{zhong2025gagrasp} \\
CGA & $\mathbb{G}_{4,1,0}$ & conformal $\supset E(3)$ & null points, circles, spheres; dilations, inversions & richest object vocabulary, distance-aware; 32 blades, indefinite metric & CGA-PoseNet~\cite{pepe2024cgaposenet}, C-GATr~\cite{de2024euclidean} \\
STA & $\mathbb{G}_{1,3,0}$ & Lorentz & events, bivectors, spinors; boosts & natural for relativistic / Maxwell systems; not for Euclidean 3D tasks & STAResNet~\cite{pepe2024staresnet}, L-GATr~\cite{spinner2024lorentz} \\
\bottomrule
\end{tabularx}
\vspace{-8pt}
\end{table*}


\textbf{Euclidean GA (EGA)} $\mathbb{G}_{3,0,0}$~\cite{miller2013geometric} has basis vectors $\{e_1, e_2, e_3\}$ and is spanned by $2^3=8$ basis blades $\{1, e_1, e_2, e_3, e_{12}, e_{13}, e_{23}, e_{123}\}$. Its multivectors contain scalars, vectors, bivectors representing oriented planes, and the pseudoscalar $e_{123}$. EGA represents rotations and reflections naturally through the $O(3)$ action, but it does not encode translations or affine subspaces natively. Tasks involving rigid motions or incidence relations among affine objects therefore often require an augmented algebra such as PGA or CGA~\cite{doran2003geometric,brehmer2023geometric,gunn2019projective}.

\textbf{Projective GA (PGA)} $\mathbb{G}_{3,0,1}$~\cite{gunn2019projective,dorst2022guided} augments 3D Euclidean directions with a null basis vector $e_0$ where ($e_0^2=0$), yielding $2^4=16$ basis blades. A vector $v = \mathbf{n} - \delta e_0$ represents an oriented plane with unit normal $\mathbf{n}$ at signed distance $\delta$ from the origin, and reflections are generated by sandwiching with such plane vectors. Rigid transformations decompose into products of plane reflections: two intersecting planes give a rotation, two parallel planes a translation~\cite{dorst2022guided}. This makes PGA well-suited for $\mathrm{E}(3)$-structured tasks~\cite{ruhe2023geometric,brehmer2023geometric}.

\textbf{Conformal GA (CGA)} $\mathbb{G}_{4,1,0}$~\cite{dorst2009geometric,doran2002conformal} embeds 3D Euclidean space into a 5D space with $2^5=32$ basis blades by introducing two additional directions, often represented through origin and infinity null vectors \(o\) and \(\infty\). A Euclidean point $p$ is embedded as a null vector $p = o + \mathbf{p} + \frac{1}{2}\mathbf{p}^2\infty$, and the inner product between embedded points encodes squared Euclidean distance~\cite{doran2002conformal}. Circles, spheres, and conformal transformations can be represented compactly as blades or versors, giving CGA the a broad geometric vocabulary. This expressiveness comes at the cost of 32-dimensional multivectors and an indefinite metric, which can introduce numerical and implementation challenges in neural architectures~\cite{dorst2022guided,de2024euclidean}.

\textbf{Spacetime Algebra (STA)} $\mathbb{G}_{1,3,0}$~\cite{hestenes2015space,10.1119/1.1571836} models Minkowski spacetime using one timelike direction and three spacelike directions, with \(e_0^2=1\) and \(e_k^2=-1\) for \(k=1,2,3\). Its multivectors encode events, spacetime bivectors, and spinorial representations of Lorentz transformations. The \emph{spacetime split} decomposes a spacetime vector into temporal and spatial components relative to a reference frame, while Lorentz transformations are represented through versor sandwich actions. STA is natural for relativistic systems, electromagnetism, and high-energy physics~\cite{dressel2015spacetime,bahns2015quantum,spinner2024lorentz}, but it is usually unnecessary for purely Euclidean 3D perception tasks.

\subsection{Representing Geometry and Transformations}
\label{sec:geometry-transformations}

The practical appeal of GA for neural-network design is that geometric objects and transformations can often be represented within the same algebraic system. The precise interpretation of a multivector depends on the chosen algebra: vectors represent directed quantities in EGA, planes in common PGA conventions, null points in CGA, and spacetime events in STA. Higher-grade blades, formed through the outer product (\(\wedge\)), represent oriented subspaces or composite geometric entities such as lines, planes, circles, and spheres, depending on the algebra. Thus, the choice of algebra fixes both the geometric vocabulary available to the model and the transformations that can be expressed naturally.

Transformations in GA are commonly represented by \emph{versors}, products of invertible vectors that act by sandwiching. For a unit vector \(n\), the map
\(x'=-nxn\)
reflects a vector \(x\) in the hyperplane orthogonal to \(n\). More general orthogonal transformations are obtained by composing such reflections. In particular, the product of two unit vectors defines an even versor, or \emph{rotor}, which represents a rotation in the plane spanned by the corresponding directions.

A rotor can be written in exponential form as
\(R = e^{-B\theta/2},\)
where \(B\) is a unit bivector specifying the rotation plane and \(\theta\) is the rotation angle. The rotated multivector is obtained by the sandwich action
\(x' = R x \tilde{R},\)
where \(\tilde{R}\) denotes reversion. This formulation is dimension-independent: rotations are associated with oriented planes rather than with axes, making the same algebraic mechanism applicable beyond three dimensions.

Rigid-body motions require both rotations and translations. In projective geometric algebra, these are represented by \emph{motors}, even versors that implement elements of \(\mathrm{SE}(3)\) through the same sandwich-action principle. Motors compactly encode rigid transformations and are therefore useful in GA-native models for poses, frames, articulated objects, and rigid-body tokens. However, motors represent orientation-preserving rigid motions; isometries involving an odd number of reflections require more general versors. For GA-native neural networks, these transformation mechanisms determine which quantities can be parameterized directly, which equivariances can be enforced by construction, and which operations require additional constraints or projections.

\subsection{Algebra Selection Guide}
\label{sec:algebra-selection}
The choice of algebra is one of the most consequential design decisions in a GA-native network, yet the literature has not converged on a single selection procedure. Table~\ref{tab:algebra_decision} provides a compact decision aid distilled from the patterns observed across the surveyed works. The primary decision variables are: (i) the geometric objects required by the task, such as positions, directions, lines, planes, circles, or spheres; (ii) the target transformation group; (iii) whether metric quantities such as distances should be represented directly in the algebra; and (iv) the available engineering budget for numerical stability, memory, and implementation complexity. The table also makes explicit when GA may be unnecessary: if the task contains no structured geometric objects, if the relevant symmetry is already handled well by a mature non-GA architecture, or if the algebraic overhead exceeds the expected inductive-bias benefit, a non-GA equivariant or conventional neural architecture (see Section~\ref{sec:non-ga-comparison}) may be preferable.

\begin{table*}[t]
\caption{Algebra-selection decision aid for GA-native neural network design. Each row corresponds to a primary task need; the recommended algebra, rationale, and avoidance conditions are synthesized from the surveyed literature.}
\label{tab:algebra_decision}
\centering
\footnotesize
\setlength{\tabcolsep}{4pt}
\renewcommand{\arraystretch}{1.15}
\begin{tabularx}{\textwidth}{@{}>{\raggedright\arraybackslash}p{3.9cm}>{\raggedright\arraybackslash}p{0.9cm}>{\raggedright\arraybackslash}X>{\raggedright\arraybackslash}p{4.3cm}>{\raggedright\arraybackslash}p{3.7cm}@{}}
\toprule
Task need & Prefer & Why & Avoid if & Representative tasks \\
\midrule
Rotation-only features on centerable data & EGA & compact (8 blades), stable, well-supported & translations matter; data cannot be centered & PDE fields, $\mathrm{O}(3)$ molecular features \\
Rigid-body tokens (positions + orientations) & PGA & native $\mathrm{E}(3)$ motors; 16 blades & distance-aware attention is critical & mesh regression, protein frames, grasps \\
Distance-sensitive or conformal geometry & CGA & squared-distance inner product; circles, spheres as blades & indefinite metric causes instability; 32-bit precision required & pose regression, distance-aware attention \\
Relativistic / Lorentz-covariant data & STA & native Lorentz group; spacetime split & task is purely Euclidean 3D & HEP jets, Maxwell surrogates \\
Only scalar / categorical features; no geometric objects & --- & GA adds overhead without representational benefit & always & tabular data, NLP, standard image classification \\
\midrule
\multicolumn{5}{@{}p{\dimexpr\textwidth-4pt}@{}}{\textit{Common failure mode if misapplied:} EGA, loses translations; PGA, cannot represent circles/spheres; CGA, unnecessary overhead for rigid geometry; STA, over-specialized for non-relativistic tasks.} \\
\bottomrule
\end{tabularx}
\vspace{-8pt}
\end{table*}

\section{GDL Principles: Equivariance and Symmetries}
Geometric Deep Learning (GDL) studies neural architectures whose inductive biases are matched to the structure and symmetries of the data domain. Following the Erlangen-inspired perspective of Bronstein et al.~\cite{bronstein2021geometric}, a learning problem is characterized by the transformations under which predictions should be invariant or equivariant. GA-native networks fit into this programme by representing geometric quantities as multivectors and using algebraic operations designed to respect the associated transformation groups.

\subsection{Leveraging Symmetries in Deep Learning}

Incorporating known symmetries into neural architectures is a central strategy in GDL. Translation equivariance in convolutional networks~\cite{lecun1995convolutional,he2016deep}, rotation equivariance in geometric and structural models~\cite{jumper2021highly}, and permutation invariance or equivariance in point-set and graph models~\cite{qi2017pointnet,zaheer2017deep} illustrate how architectural constraints can improve data efficiency, generalization under transformations, and robustness~\cite{bronstein2021geometric}. GA-native layers pursue the same goal through algebraic structure: multivector representations, geometric products, grade projections, and versor actions can be arranged so that features transform consistently under the target symmetry group~\cite{brehmer2023geometric,ruhe2023clifford,spinner2024lorentz,liu2024clifford,zhdanov2024clifford,tran2024clifford,pepe2024clifford,liu2024multivector}. In favorable settings, this can reduce reliance on data augmentation and improve sample efficiency~\cite{ruhe2023geometric,wagner2024generating,brehmer2023geometric,zhong2025gagrasp}. A controlled GATr scaling study~\cite{brehmer2025does} finds that augmentation can close the data-efficiency gap given sufficient epochs, while GATr outperforms non-equivariant models at every tested compute budget.

\subsection{Foundational Group Theory for Equivariance}

We briefly recall the group-theoretic notation needed for the Clifford-group discussion below. 
A \emph{group} \((G,\circ)\) is a set equipped with a binary operation satisfying closure, associativity, identity, and inverse axioms. Important geometric groups include the orthogonal group \(\mathrm{O}(n)\), consisting of origin-preserving isometries such as rotations and reflections; the special orthogonal group \(\mathrm{SO}(n)\subset \mathrm{O}(n)\), consisting of orientation-preserving rotations; and the special Euclidean group \(\mathrm{SE}(n)\), consisting of orientation-preserving rigid motions, i.e., rotations and translations. A subgroup \(H\subseteq G\) is a subset that is itself a group under the same operation.

A \emph{group action} \(G\times X\to X\) specifies how each group element transforms elements of a set \(X\), satisfying \(e\cdot x=x\) and \((g\circ h)\cdot x=g\cdot(h\cdot x)\). When \(X\) is a vector space \(V\), a linear action is often represented by a homomorphism \(\rho:G\to GL(V)\), so that \(\rho(g_1\circ g_2)=\rho(g_1)\rho(g_2)\). In neural networks, \(\rho_X\) and \(\rho_Y\) denote how the group acts on input and output feature spaces, respectively.

\begin{figure}[t]
\centering
\includegraphics[width=\columnwidth]{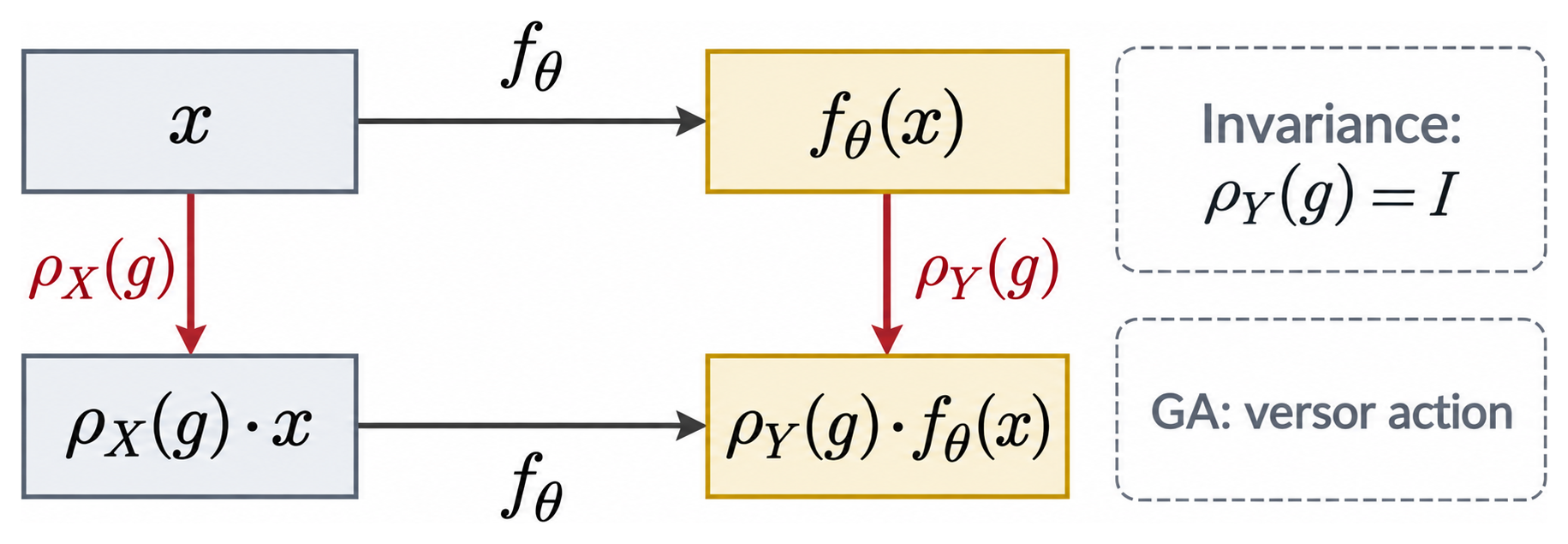}
\caption{Commutative view of equivariance. A NN layer is equivariant when transforming the input and then applying $f_\theta$ yields the same result as applying $f_\theta$ first and then transforming the output. Invariance is the special case with trivial output action.}
\label{fig:equivariance-diagram}
\vspace{-8pt}
\end{figure}

A map \(f:X\to Y\) is \(G\)-\emph{equivariant} if
\(f(\rho_X(g)x)=\rho_Y(g)f(x)\)
for all \(g\in G\) and \(x\in X\). Equivariance means that transforming the input and then applying \(f\) gives the same result as applying \(f\) first and then transforming the output. The special case in which the output action is trivial, \(\rho_Y(g)=I\), gives \(G\)-\emph{invariance}: \(f(\rho_X(g)x)=f(x)\). Fig.~\ref{fig:equivariance-diagram} gives the standard commutative view.

\subsection{Achieving Equivariance with Geometric Algebra}
\label{sec:achieving-equivariance-ga}
GA provides a structured route to equivariant neural-network design because many geometric transformations can be represented as versor or Clifford-group actions on multivectors. When layers are built from operations that commute with these actions, such as geometric products, grade projections, invariant scalar functions, and appropriate sandwich actions, the resulting maps can be equivariant by construction~\cite{de2024euclidean,wagner2024generating,spinner2024lorentz,brehmer2023geometric,zhong2025gagrasp}.

Different GAs support different transformation groups and therefore induce different geometric biases. In Euclidean GA \(\mathbb{G}_{n,0,0}\), rotations and reflections are represented through the associated \(\mathrm{Spin}(n)\) and \(\mathrm{Pin}(n)\) actions, which double-cover \(\mathrm{SO}(n)\) and \(\mathrm{O}(n)\), respectively~\cite{gallier2008clifford,ruhe2023geometric}. Operations such as geometric products, grade projections, and rotor actions can preserve \(\mathrm{SO}(n)\)- or \(\mathrm{O}(n)\)-covariance when used in architectures that respect these actions~\cite{geiger2022e3nn}.

Projective GA \(\mathbb{G}_{n,0,1}\) is well suited to Euclidean equivariance. Even versors, known as \emph{motors}, implement orientation-preserving rigid motions from the \(\mathrm{SE}(n)\) group, and odd versors, including reflections, implement motions from the larger \(\mathrm{E}(n)\) group. These transformations act on PGA representations of points, lines, and planes through sandwich products. The Geometric Algebra Transformer (GATr)~\cite{brehmer2023geometric} and related models~\cite{wagner2024generating,de2024euclidean,zhong2025gagrasp,suk2024lab,suk2024geometric,sun2025hybrid} are among the architectures that use PGA-style multivectors for \(\mathrm{E}(3)\)- or \(\mathrm{SE}(3)\)-structured learning. Conformal GA \(\mathbb{G}_{n+1,1,0}\) can also represent Euclidean transformations as a subgroup of conformal transformations~\cite{de2024euclidean}.

Spacetime Algebra (STA) $\mathbb{G}_{1,3,0}$ is the natural algebraic setting for Lorentz-covariant modeling. Neural networks built using STA or Lorentzian GA variants, such as L-GATr~\cite{spinner2024lorentz,brehmer2024lorentz}, can be made equivariant to Lorentz transformations, which is central in relativistic physics.

This hierarchy means that the choice of GA strongly influences the symmetry group that a network can represent naturally. Choosing an algebra that includes more structure than the task requires may introduce unnecessary dimensionality, constraints, or numerical complexity~\cite{brehmer2023geometric}, whereas choosing an algebra with insufficient structure may leave important geometric relationships outside the model's native representation. GA-native models should therefore be understood as one route to geometry-aware learning, alongside non-GA equivariant approaches such as group-equivariant CNNs~\cite{cohen2016group}, steerable CNNs~\cite{cohen2016steerable}, Tensor Field Networks~\cite{thomas2018tensor}, and \(\mathrm{E}(n)\)-equivariant GNNs~\cite{satorras2021n}. The distinctive appeal of GA is not equivariance alone, which these alternatives also provide, but the possibility of expressing heterogeneous geometric entities and their transformations within a single algebraic framework, potentially simplifying layers that must reason jointly over positions, orientations, and higher-order objects.

\subsection{Comparison with Non-GA Equivariant Approaches}
\label{sec:non-ga-comparison}
GA-native networks are one of several routes to geometry-aware learning. To help practitioners decide when GA is likely to be useful, Table~\ref{tab:non_ga_comparison} compares the principal equivariant paradigms along dimensions that matter for architecture selection. Three high-level takeaways emerge from this comparison:

\begin{enumerate}[leftmargin=*,itemsep=1pt]
\item \textbf{Standard \(\mathrm{E}(3)\) tasks with scalar/vector point features.} When the data consists of point clouds or graphs with primarily scalar and vector attributes and the target symmetry is \(\mathrm{E}(3)\), mature non-GA methods, particularly \(\mathrm{E}(n)\)-equivariant GNNs~\cite{satorras2021n} and steerable or tensor-field architectures~\cite{cohen2016steerable,thomas2018tensor}, are strong defaults with well-optimized software stacks. GA adds value primarily when the task also involves heterogeneous geometric entities, such as lines, planes, frames, or rigid motions, or when a unified algebra simplifies layer design.
\item \textbf{Mixed geometric objects and frames.} Tasks that require joint reasoning over positions, orientations, higher-order subspaces, or coupled physical fields are where GA is most compelling. Representing points, lines, planes, motors, and their products within one algebraic framework can reduce the need for bespoke per-entity layers.
\item \textbf{Spacetime and conformal symmetries.} For Lorentz-covariant or conformally structured tasks, STA and CGA provide native algebraic support for the transformations and geometric objects. Non-GA alternatives may still be possible, but often require more specialized representation theory or task-specific engineering. For instance, Lorentz local canonicalization (LLoCa)~\cite{spinner2025lorentz} makes arbitrary backbones Lorentz-equivariant via equivariantly predicted local frames and reports lower training cost than L-GATr.
\end{enumerate}
\noindent GA-native methods should not be interpreted as dominating other geometric learning paradigms. In many settings, group-equivariant, steerable, tensor-field, or $\mathrm{E}(n)$-equivariant architectures provide lower-complexity and better-validated choices, especially when the task symmetry is known and can be captured efficiently without multivector representations.

\begin{table*}[t]
\caption{Comparison of equivariant paradigms for geometric learning. GA-native networks are most attractive when heterogeneous geometric entities or non-standard symmetry groups are central to the task; mature non-GA methods often remain preferable when the symmetry and data type are already well matched.}
\label{tab:non_ga_comparison}
\centering
\footnotesize
\setlength{\tabcolsep}{3.5pt}
\renewcommand{\arraystretch}{1.15}
\begin{tabularx}{\textwidth}{@{}>{\hsize=0.8\hsize\raggedright\arraybackslash}X>{\hsize=0.7\hsize\raggedright\arraybackslash}X>{\hsize=0.9\hsize\raggedright\arraybackslash}X>{\raggedright\arraybackslash}p{1.25cm}>{\raggedright\arraybackslash}p{1.0cm}>{\raggedright\arraybackslash}p{1.45cm}>{\hsize=1.37\hsize\raggedright\arraybackslash}X>{\hsize=1.23\hsize\raggedright\arraybackslash}X@{}}
\toprule
Paradigm & Latent repr. & Typical symmetry & Mixed entities? & Join / meet? & Software maturity & Best-fit regime & Limitations \\
\midrule
Group-equiv.\ CNNs~\cite{cohen2016group} & feature maps on group & discrete / compact & No & No & High & grid data with known discrete symmetry & fixed grid; limited to compact groups \\
Steerable CNNs~\cite{cohen2016steerable} & irrep.\ fiber bundles & $SO(n)$, $O(n)$, $E(n)$ & Partial & No & High & continuous rotation equiv.\ on grids & Clebsch--Gordan overhead; grid-based \\
TFN / SE(3)-Transf.~\cite{thomas2018tensor} & spherical tensors & $SE(3)$ & Partial & No & Medium & point clouds, molecules & high-order tensor products expensive \\
$E(n)$-equiv.\ GNNs~\cite{satorras2021n} & scalar + coord.\ vectors & $E(n)$ & No & No & High & molecular graphs, $n$-body & limited to scalar \& vector features \\
Lie-group / pose nets & Lie algebra elements & $SE(3)$, $SO(3)$ & Partial & No & Medium & pose estimation, SLAM & task-specific parametrizations \\
GA-native networks & multivectors & $O(n)$, $E(n)$, conformal, Lorentz & Yes & Yes & Low--Med & mixed objects, coupled fields, non-standard symmetries & dim.\ overhead; numerical instability in CGA/STA \\
\bottomrule
\end{tabularx}
\vspace{-8pt}
\end{table*}

\subsection{The Clifford Group and its Action in Equivariant Networks}

A more formal route to GA-native equivariance is through the \emph{Clifford group}, also called the Lipschitz or Clifford--Lipschitz group~\cite{vaz2016introduction,lounesto2001clifford}. Let \(V\) be a finite-dimensional real vector space equipped with a non-degenerate quadratic form \(q\), and let \(Cl(V,q)\) be its Clifford algebra. The Clifford group \(\Gamma(V,q)\) is the subgroup of invertible algebra elements whose twisted adjoint action on the generating vector space preserves (V) and acts as an orthogonal transformation with respect to (q). In common conventions, this action takes the form \(
x \mapsto g x g^{-1}
\quad \text{or} \quad
x \mapsto \alpha(g)xg^{-1}, \)
where \(\alpha\) denotes the grade involution~\cite{ruhe2023clifford,liu2024clifford,ali2024metric,maruyama2024gradient}. The Pin and Spin groups arise as subgroups corresponding to orthogonal and orientation-preserving orthogonal transformations, respectively.

For neural networks, the significance of the Clifford-group action is that it extends consistently from vectors to the full Clifford algebra. An orthogonal transformation of \(V\) induces a grade-preserving algebra action on multivectors, so scalars, vectors, bivectors, and higher-grade components transform according to compatible rules~\cite{ruhe2023clifford}. This is essential for GA-native networks, which often operate on full multivectors rather than vector features alone. Without such an extended action, equivariance on the vector subspace would not determine how higher-grade components should transform, weakening the geometric interpretation of learned multivector features.

CGENNs~\cite{ruhe2023clifford} exploit this structure by constructing layers from operations that commute with the Clifford-group action. In particular, geometric products, grade projections, scalar linear combinations, and invariant scalar nonlinearities can be composed to form equivariant multivector maps. This provides a principled parameterization strategy: rather than learning arbitrary coordinate-wise maps on multivectors, the architecture restricts learnable functions to combinations of Clifford-compatible operations. Subsequent work extends or adapts this idea in several directions. Ali et al.~\cite{ali2024metric} learn the underlying quadratic form while preserving Clifford-equivariant structure; Liu et al.~\cite{liu2024clifford} use related polynomial constructions in simplicial architectures; Brandstetter et al.~\cite{brandstetter2022clifford} apply Clifford layers to PDE modeling; and Liu et al.~\cite{liu2024multivector} implement grade-structured polynomial multivector functions for \(\mathrm{O}(n)\)-equivariant architectures. Filimoshina and Shirokov~\cite{filimoshina2025glgenn} build parameter-light GLGENNs on generalized Lipschitz groups that also cover degenerate algebras such as PGA, sharing weights across grade subspaces. Across these examples, the common design principle is that equivariance follows from the closure of the chosen layer primitives under the Clifford-group action, not from arbitrary multivector-valued parameterization.

\section{GA NN Layers}
GA-native neural networks require layer primitives that operate directly on multivector-valued features while preserving the intended geometric transformation laws. This section reviews the main building blocks used in such architectures: multivector embeddings, GA-parametrized linear and bilinear maps, transformation-based layers, equivariant activations and normalizations, and GA-based attention mechanisms.

\subsection{Multivector Representations}
GA-native architectures begin by embedding data into a chosen geometric algebra. Scalars, vectors, and higher-order geometric objects are assigned to algebra-specific components, such as grade-0 scalars, grade-1 vectors, PGA points or motors, CGA null points, or bivectors. These multivectors then serve as the feature channels processed by the network. Depending on the architecture, grades may be preserved, mixed through geometric products, gated independently, or constrained to maintain equivariance~\cite{liu2024clifford,liu2024multivector}. Thus, the algebra choice determines both the initial encoding and the layer operations, symmetry constraints, and computational overhead available to the model~\cite{de2024euclidean}.

\subsubsection{Embedding Primitive Data}

The simplest embeddings map primitive feature types to algebraic components with matching transformation behavior. Scalar quantities, such as mass, charge, time-independent material parameters, or categorical labels, are invariant under spatial transformations and are therefore represented as grade-0 elements in $Cl^{(0)}(V,q)$~\cite{ruhe2023clifford,brehmer2023geometric}. Directional or vector-valued quantities, such as velocity, force, displacement, or local orientation features, transform covariantly under rotations and reflections and are commonly embedded as grade-1 elements in $Cl^{(1)}(V,q)$~\cite{ruhe2023clifford,brehmer2023geometric}. Position coordinates require more care: they may be represented as ordinary vectors in EGA, homogeneous objects in PGA, or null-point embeddings in CGA, depending on the target symmetry and algebra. 

\subsubsection{Representing Complex Geometric Structures}
More structured inputs require embeddings that reflect both the task geometry and the chosen algebra. Point clouds and meshes illustrate the main trade-offs summarized in Table~\ref{tab:ga_encodings}. In EGA ($\mathbb{G}_{n,0,0}$), a point is represented by an origin-relative position vector, giving natural $O(n)$ or $SO(n)$ transformation behavior but requiring centering or additional mechanisms for translation equivariance~\cite{de2024euclidean}. PGA ($\mathbb{G}_{n,0,1}$) introduces a null basis vector $e_0$ and represents Euclidean incidence geometry more directly: in 3D PGA, points are typically represented as trivectors, lines as bivectors, and planes as vectors. This makes PGA a natural choice for architectures such as GATr~\cite{brehmer2023geometric}, which process point clouds or meshes without manual centering. CGA ($\mathbb{G}_{n+1,1,0}$) embeds Euclidean space into a higher-dimensional algebra in which null-point inner products encode squared Euclidean distance up to convention-dependent sign and scale~\cite{sveier2017object}. It represents circles and spheres as blades, at the cost of higher dimensionality, e.g., $2^5=32$ basis blades in 3D, and possible numerical issues from the indefinite metric~\cite{de2024euclidean}.

For graph-structured data, node, edge, or simplex features can be represented as multivectors~\cite{zhong2025graph}. Recent simplicial models~\cite{liu2024clifford} initialize higher-order relationships algebraically: an edge $(v_i,v_j)$ may be represented through the geometric product $h_{v_i}h_{v_j}$, while a triangle may be initialized as $h_{v_i}h_{v_j}h_{v_k}$. These constructions encode higher-order interactions through GA operations rather than treating edges and simplices as unrelated feature channels.

GA representations are also useful for coupled physical fields. In computational fluid dynamics, a scalar pressure field $p(x)$ and vector velocity field $u(x)$ can be represented as a single multivector field $M(x)=p(x)+u(x)$, allowing Clifford convolutional layers to process coupled scalar-vector quantities jointly~\cite{brandstetter2022clifford}. Similarly, in electromagnetic settings, the electric field $\mathbf{E}$ and magnetic field $\mathbf{B}$ can be combined into a multivector field such as $F=\mathbf{E}+I\mathbf{B}$, where $I$ is the pseudoscalar of the algebra. These examples illustrate how multivector encodings can align network channels with physically or geometrically coupled quantities.

Table~\ref{tab:ga_encodings} summarizes encoding choices for network design.

\begin{table*}[t]
\caption{Recurring object encodings in GA-native neural networks. The entries summarize common conventions; exact representations may vary with algebra conventions and implementation choices.}
\label{tab:ga_encodings}
\centering
\footnotesize
\setlength{\tabcolsep}{4pt}
\renewcommand{\arraystretch}{1.15}
\begin{tabularx}{\textwidth}{@{}l>{\raggedright\arraybackslash\hsize=.95\hsize}X>{\raggedright\arraybackslash\hsize=.85\hsize}X>{\raggedright\arraybackslash\hsize=.7\hsize}X>{\raggedright\arraybackslash\hsize=.8\hsize}X>{\raggedright\arraybackslash\hsize=1.2\hsize}X>{\raggedright\arraybackslash\hsize=1.5\hsize}X@{}}
\toprule
Algebra & Point & Affine objects & Motion & Distance cue & Best-fit tasks & Representative works \\
\midrule
EGA & position vector (origin-relative) & origin-passing subspaces & rotors & Euclidean norm & rotation-aware features, PDE fields & CGENN~\cite{ruhe2023clifford}, Clifford layers~\cite{brandstetter2022clifford} \\
PGA & homog.\ trivector & planes, lines, points & motors & not directly distance-aware & $SE(3)$ tokens, meshes, poses & GATr~\cite{brehmer2023geometric}, GAFL~\cite{wagner2024generating} \\
CGA & null-point embedding & lines, circles, spheres & versors / motors & squared-dist.\ inner product & distance-sensitive attention, richer geometry & iP-GATr / C-GATr~\cite{de2024euclidean}, CGA-PoseNet~\cite{pepe2024cgaposenet} \\
STA & event / 4-vector & spacetime planes & rotors / boosts & Minkowski interval & relativistic physics, HEP & L-GATr~\cite{spinner2024lorentz}, STAResNet~\cite{pepe2024staresnet} \\
\bottomrule
\end{tabularx}
\vspace{-8pt}
\end{table*}

\subsection{GA-Parametrized Linear and Bilinear Layers}
Standard linear layers in NNs ($y=Wx+b$) must be constrained when inputs and outputs are multivectors and equivariance is required. CGENN~\cite{ruhe2023clifford} shows that a Clifford-group-equivariant linear map $\phi(V,q)\rightarrow Cl(V,q)$ can be parameterized by learnable scalar weights applied independently to grade projections:
\begin{equation}
    \phi (x) = \sum_k w_k \langle x \rangle_k ,
\end{equation}
where $\langle x \rangle_k$ denotes the grade-$k$ component of $x$ and $w_k \in \mathbb{R}$ are learnable parameters. Because each grade subspace transforms consistently under the Clifford-group action, this grade-wise parameterization preserves equivariance while allowing the network to rescale different geometric types. Later architectures, such as GATr~\cite{brehmer2023geometric}, extend this idea by allowing selected equivariant maps between grades when mediated by algebraically meaningful elements. In PGA, for example, multiplication by the null basis vector $e_0$ can be used to construct additional equivariant channels, yielding maps of the form $\phi (x) = \sum_k w_k \langle x \rangle_k + \sum_k v_k e_0 \langle x \rangle_k$.

Linear grade-wise maps alone provide limited interaction among geometric components. Bilinear layers address this limitation by using GA products to mix multivector channels while preserving the target symmetry when the product is compatible with the group action. Geometric-product layers, used in models such as CGENN, GATr, and related architectures~\cite{de2024euclidean,wagner2024generating,brehmer2023geometric,ruhe2023clifford}, compute products between input multivectors or multivector channels. A fully connected geometric-product layer~\cite{ruhe2023clifford} takes multivector channels $x_i$ and $y_i$ and computes
\begin{equation}
    z_j = \sum_i P_{\phi_{ij}} (x_i, y_i),
\end{equation}
where $P_\phi$ denotes a parameterized product with learnable scalar weights for allowed grade interactions. Such layers can generate scalar, vector, bivector, and higher-grade components from lower-grade inputs, enabling cross-grade feature construction that is unavailable to purely grade-wise linear maps.

Other algebraic bilinear operations can be useful in algebras with incidence structure. In PGA, for instance, join and meet operations can be implemented through combinations of duality and the outer product, such as dualized products of the form $(x^\ast\wedge y^\ast)^\ast$, depending on the convention. These operations allow networks to construct or compare geometric objects such as lines through points, planes through points, or intersections of subspaces. In architectures such as GATr and its variants~\cite{brehmer2023geometric,de2024euclidean,wagner2024generating}, such incidence-aware operations provide additional equivariant pathways for mixing information beyond the geometric product alone.

\subsection{Transformation-based Layers}
The layers above learn equivariant feature representations. A complementary design strategy is to parameterize geometric transformations directly, so that some weights correspond to versors, rotors, motors, or other algebraic operators. This can provide a more structured inductive bias when the task involves rotations, rigid motions, pose updates, or transformation composition, although the interpretability depends on maintaining the algebraic constraints that make these parameters valid geometric transformations.

In GA, rotations and rigid motions can be represented by versors acting on multivectors through sandwich products, $x' = MxM^{-1}$, with $M^{-1} = \tilde{M}$ for normalized rotors or motors. CGA-PoseNet~\cite{pepe2024cgaposenet} uses this idea by designing layers where the inputs, outputs, weights and biases are represented as motors. A dense layer in this framework computes:
\begin{equation}
    M_{\mathrm{out}} = \sum_{i=1}^{C_{\mathrm{in}}} W_i M_i \tilde{W}_i + B_i ,
\end{equation}
where each learned motor $W_i$ applies a rigid-motion-like transformation to the input motor $M_i$. The layer can therefore be interpreted as aggregating transformed motor-valued features, provided the learned quantities are constrained or normalized to remain valid motors.

More generally, GCAN~\cite{ruhe2023geometric} introduces group-action layers, that learn a combinations of prescribed group actions: $T(x) = \sum_i w_i \cdot a_i x \tilde{a}_i$, where $a_i$ are learnable versors or group elements and $w_i$ are scalar weights. These layers can be viewed as learnable mixtures of geometric actions, offering an explicit transformation-based alternative to purely coordinate-wise or grade-wise feature mixing. Beyond geometric feature processing, Pence et al.~\cite{pence2025composing} use compositions of rotors to construct structured approximations of linear transformations. Their approach requires only $O(\log^2 d)$ parameters, compared with $O(d^2)$ for dense linear maps, and achieves competitive performance when applied to attention projections in large language models. This illustrates how GA-based transformation parameterizations can also improve parameter efficiency in architectures without explicit geometric inputs.

\subsection{Activations and Normalizations}
Introducing non-linearity and ensuring stable training while preserving equivariance are central challenges in GA NN design. Standard scalar functions applied element-wise to a multivector's coefficients will break the geometric structure.

In terms of activation functions, the most common and principled solution is gating, where a multivector is scaled by an invariant scalar computed from itself. For a multivector $x$, the activation is of the form $\psi (x) = \sigma (\phi (x)) x$, where $\phi (x)$ is an invariant function. MSiLU in GCANs~\cite{ruhe2023geometric} applies a sigmoid gate to each $k$-vector component of a multivector independently: $\sigma(f_k (x)) \cdot \langle x \rangle_k$, where $f_k(x)$ is a learnable scalar function of $x$. Similarly, in CGENNs~\cite{ruhe2023clifford}, this takes the form $\sigma (\lVert \langle x \rangle_k \rVert^2 ) \langle x \rangle_k$ for each non-scalar grade $k$, where the norm is invariant under the group action. Another common strategy~\cite{brehmer2023geometric,brehmer2024lorentz,zhong2025gagrasp} is to use the scalar part of a multivector to gate the entire multivector. GATr~\cite{brehmer2023geometric} employs a scalar-gated GELU~\cite{hendrycks2016gaussian}: $\mathrm{GELU}(\langle x \rangle_0) x$, which preserves equivariance if the scalar part is invariant and the multiplication is the GA geometric product or scalar multiplication. More advanced, algebra-aware activations have been proposed, such as the GA-ReLU~\cite{pepe2024ga}. For 2D CFD problems, GA-ReLU decomposes a multivector into its "spinor" and "vector" parts (which are isomorphic to two complex numbers) and applies a phase-dependent cardioid activation. This models interactions between vector components in a geometrically meaningful way that a simple norm-based gating would miss. Recently, the Kolmogorov-Arnold representation theorem~\cite{liu2024kan} has been adapted to multivector spaces via Clifford Kolmogorov-Arnold Networks (ClKAN)~\cite{wolff2026clifford}. ClKAN proves that KANs can be natively implemented over Clifford algebras, utilizing Randomized Quasi Monte Carlo grid generation to manage high-dimensional scaling, thus offering a completely new paradigm for geometric non-linear function approximation.

Standard techniques like BatchNorm~\cite{ioffe2015batch} are not equivariant. GA-NNs require normalization schemes that respect the algebraic structure. CGENNs~\cite{ruhe2023clifford} and GCANs~\cite{ruhe2023geometric} use variants of GroupNorm~\cite{wu2018group} or custom norm-based schemes that normalize each $k$-vector component of a multivector across channels, again preserving the grade structure by only involving subtraction of an average $k$-vector and scalar division/multiplication. GATr~\cite{brehmer2023geometric} employs a LayerNorm~\cite{ba2016layer} defined using the GA inner product: $x/\sqrt{\mathbb{E}_c [\langle x \cdot x \rangle_0]}$, which is invariant. This approach faces challenges in GAs with indefinite metrics (like CGA or STA), where $x\cdot x$ can be zero or negative. Adaptations~\cite{brehmer2024lorentz,de2024euclidean}, such as using the absolute value of the inner product for each grade or careful numerical handling, are required.

\subsection{GA-based Attention Mechanisms}
With the success of Transformer architectures~\cite{vaswani2017attention,dosovitskiy2020image}, attention mechanisms have been adapted for GA NNs. The standard scaled dot-product attention relies on the vector dot product to compute a scalar similarity score between query and key vectors. The GA inner product, $\langle q \cdot k \rangle_0$, is the direct, geometrically principled replacement. Since the inner product between two multivectors is invariant to the action of the corresponding Clifford group, the resulting attention scores are guaranteed to be invariant~\cite{brehmer2023geometric,de2024euclidean}. The value vectors $v$ can remain as full multivectors, and the output is an attention-weighted sum of these multivectors. The result is a multivector that transforms equivariantly, making this the core mechanism in~\cite{de2024euclidean,brehmer2023geometric,brehmer2024lorentz}. Clifford Frame Attention (CFA)~\cite{wagner2024generating} extends the Invariant Point Attention (IPA) from AlphaFold2~\cite{jumper2021highly} for protein design. CFA uses multivectors in PGA to represent residue frames and geometric features. Messages between residues are constructed using geometric bilinear layers operating on these multivector features, and attention weights modulate these interactions. Geometric Algebra Attention Networks~\cite{spellings2021geometric} compute attention scores from the geometric products of all input vector tuples (pairs, triplets, etc.). It achieves permutation equivariance and rotation invariance by summing over all permutations and using invariant components of the geometric products. A key limitation of the PGA inner product is that it is not sensitive to the Euclidean distance between the points represented by the multivectors. To address this, iP-GATr and C-GATr~\cite{de2024euclidean} implement distance-aware attention by mapping the point-representing multivectors from PGA to CGA and using this CGA inner product to compute attention scores. This allows the attention mechanism to directly incorporate relative positions, which is a crucial inductive bias in many physical systems.

\begin{figure*}[t]
\centering
\includegraphics[width=0.95\textwidth]{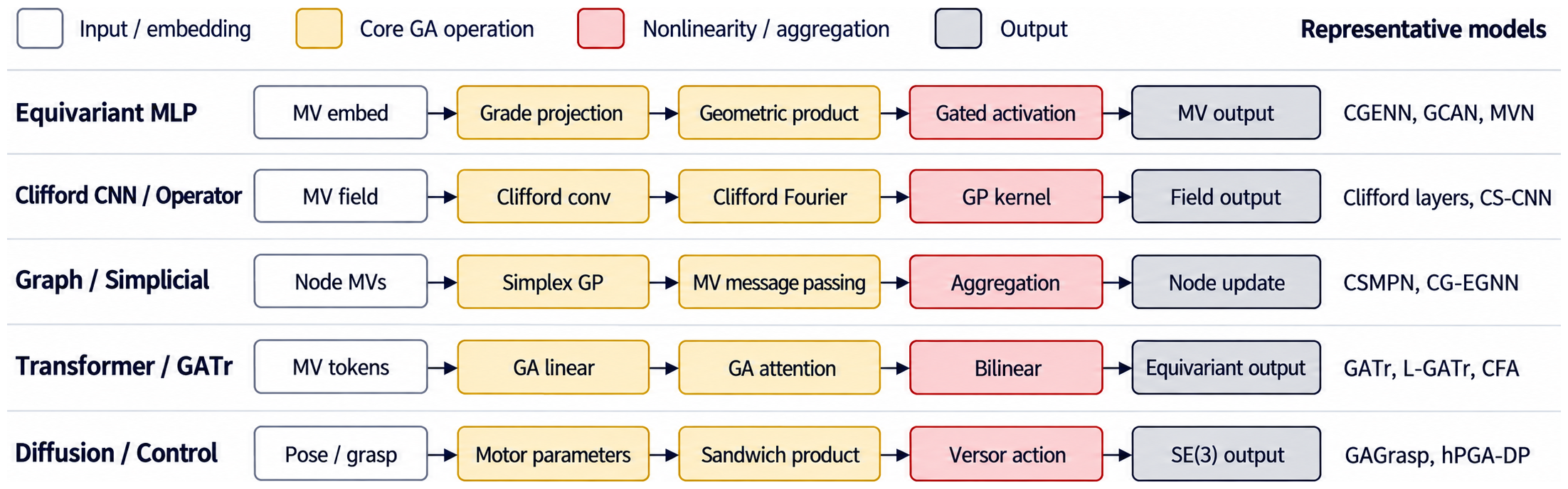}
\caption{Canonical design motifs in GA-native architectures. Different model families reuse a shared set of GA primitives but combine them in distinct computational patterns.}
\label{fig:ga-architecture-motifs}
\vspace{-8pt}
\end{figure*}

\subsection{Building GA-NN Architectures}
Building on the layer primitives above, researchers have developed a variety of complete GA-NN architectures. The earliest attempts to integrate Clifford algebras into neural networks viewed them as generalizations of complex-valued or quaternion-valued networks~\cite{hirose2006complex,rahman2001multiexpert,pearson2003clifford,yi2006learning,buchholz2008polarized,li2019an}. Pearson and Bisset~\cite{pearson1994neural} proposed Clifford MLPs for pattern recognition, and Buchholz and Sommer~\cite{buchholz2008clifford,buchholz2001clifford,Buchholz2001} further investigated Clifford neurons and learning rules. These pioneering works established the concept of multivector-valued neurons but focused on lower-dimensional algebras without fully exploiting equivariance. Modern architectures~\cite{ruhe2023clifford,liu2024multivector,kolyvakis2025geometric,pepe2024clifford} build on this foundation with principled equivariance guarantees, forming the backbone of more complex models.

As discussed in earlier sections, Geometric Clifford Algebra Networks (GCANs)~\cite{ruhe2023geometric} represent a significant step towards leveraging the geometric power of GA in a principled way. A key design principle in GCANs is that layers ensure $k$-vectors map to $k$-vectors, preserving the geometric type of features. GCANs have been successfully applied to modeling dynamical systems, including 3D rigid body transformations and fluid dynamics simulations (e.g., shallow water equations, Navier-Stokes equations, often using $\mathbb{G}_{3,0,0}$ for these examples).

For data on regular grids, such as images or volumetric fields from PDE simulations, the convolution operation can be generalized. Clifford Convolutions~\cite{brandstetter2022clifford} replace scalar-valued kernels with multivector-valued kernels that are convolved with multivector feature fields using the geometric product. GA-CNNs~\cite{li2022ga} further proposes representing different spectral bands of a hyperspectral image (HSI) as components of a GA multivector. Similarly, Clifford Fourier Layers~\cite{brandstetter2022clifford,roy2024clifford,alfarraj2022geometric} replace the standard Fast Fourier Transform (FFT) with a Clifford-Fourier Transform, enabling spectral methods on multivector fields. Clifford-Steerable CNNs~\cite{zhdanov2024clifford} achieve equivariance to groups like Pseudo-Euclidean groups $\mathrm{E}(p,q)$ by parameterizing the steerable convolution kernels implicitly with a CGENN-style MLP, which learns the kernel's values as an equivariant function of the relative position vector. Conditional CS-CNNs (C-CSCNN)~\cite{szarvas2026conditional} show that this kernel basis is incomplete and improve expressivity by conditioning kernels on equivariant representations of the input field.

GA primitives are readily integrated into the message-passing framework of GNNs. In models like~\cite{liu2024clifford,tran2024clifford,zhong2025graph}, the message and update functions are implemented as GA-NN blocks (e.g., CGENNs~\cite{ruhe2023clifford}) or GA-based primitives (e.g., geometric products). Messages, aggregated information, and node/edge features are all represented as multivectors. This allows the network to pass and process rich geometric information, including higher-order simplex features derived from geometric products, across the graph structure while maintaining full equivariance.

The Transformer architecture~\cite{vaswani2017attention}, with its attention mechanism, has revolutionized sequence processing and is increasingly applied to other data modalities~\cite{dosovitskiy2020image}. The Geometric Algebra Transformer (GATr) family~\cite{brehmer2023geometric,de2024euclidean,brehmer2024lorentz,spinner2024lorentz,suk2025vine} adapts this powerful paradigm for geometric data using GA. This family represents a significant advancement by combining the scalability and expressiveness of Transformers with the principled geometric equivariance provided by GA. The original GATr~\cite{brehmer2023geometric} is designed to be a general-purpose $\mathrm{E}(3)$-equivariant architecture. It utilizes Projective Geometric Algebra ($\mathbb{G}_{3,0,1}$) to represent inputs, hidden states, and outputs as 16-dimensional multivectors. Follow-up work~\cite{de2024euclidean} explored variants of GATr using EGA (E-GATr), a simpler PGA (P-GATr), an improved PGA (iP-GATr, similar to the original GATr), and CGA (C-GATr). This study highlighted the trade-offs between these algebras in terms of expressivity, the size of the symmetry group effectively handled, sample efficiency, and training stability. ViNE-GATr~\cite{suk2025vine} addresses the scalability of $\mathrm{E}(3)$-equivariant transformers by projecting dense geometric data into a coarse-grained latent space of virtual nodes. L-GATr~\cite{brehmer2024lorentz} extends the GATr framework to achieve Lorentz equivariance by employing Spacetime Algebra ($\mathbb{G}_{1,3,0}$). It introduces new equivariant layers (linear maps, attention, LayerNorm) specifically designed for STA. L-GATr has shown state-of-the-art results in high-energy physics applications, including amplitude regression, jet classification, and as the basis for Lorentz-equivariant generative models~\cite{spinner2024lorentz}. In core sequence modeling, Versor~\cite{huy2026versor} replaces traditional linear vector sequence operations with CGA operations to inherently respect the symmetries of the physical universe, combating the "Euclidean Bottleneck".

Beyond these general architectural classes, GA is also being incorporated into models tailored for specific, complex geometric tasks, such as protein design~\cite{wagner2024generating}, implicit neural volumes~\cite{sivgin2024geometric}, and grasp generation~\cite{zhong2025gagrasp}. Table~\ref{tab:architecture_taxonomy} summarizes the main family structure. These specialized architectures highlight a trend towards deeply integrating the specific geometric algebra and its elements that best describe the intrinsic geometry and symmetries of the target problem domain. This tailored approach often yields significant benefits in performance and data efficiency. We will discuss these tailored methods and applications in the following sections.

\begin{table*}[t]
\caption{Taxonomy of GA-native neural network families, with representative cited works.}
\label{tab:architecture_taxonomy}
\centering
\footnotesize
\setlength{\tabcolsep}{3.5pt}
\renewcommand{\arraystretch}{1.15}
\begin{tabularx}{\textwidth}{@{}>{\raggedright\arraybackslash\hsize=.8\hsize}X>{\raggedright\arraybackslash\hsize=1.25\hsize}X>{\raggedright\arraybackslash\hsize=1.0\hsize}X>{\raggedright\arraybackslash\hsize=.7\hsize}X>{\raggedright\arraybackslash\hsize=1.45\hsize}X>{\raggedright\arraybackslash\hsize=.8\hsize}X@{}}
\toprule
Family & Core GA primitive & Typical data & Symmetry & Representative works & Main caveat \\
\midrule
Early Clifford MLPs & multivector linear maps, adapted activations & low-dim.\ signals & limited & Clifford MLPs~\cite{pearson2003clifford,buchholz2008clifford} & weak equivariance framing \\
Equivariant blocks & grade projections, geometric products, gating & tokens, sets, graphs & $O(n)$, $E(n)$ & CGENN~\cite{ruhe2023clifford}, GCAN~\cite{ruhe2023geometric}, MVN/MVP~\cite{liu2024multivector}, GLGENN~\cite{filimoshina2025glgenn} & cost grows with algebra dim. \\
Clifford CNN / operator & Clifford convolution, Fourier layers, GP kernels & grids, fields, PDE states & translation + geometric & Clifford layers~\cite{brandstetter2022clifford}, CS-CNN~\cite{zhdanov2024clifford}, Fengbo~\cite{pepefengbo}, C-CSCNN~\cite{szarvas2026conditional} & specialized kernels \\
Graph / simplicial & multivector message passing, simplex GP features & graphs, molecules, simplicial complexes & perm.\ + $O(n)$/$E(n)$ & CSMPN~\cite{liu2024clifford}, CG-EGNN~\cite{tran2024clifford}, GGAN~\cite{zhong2025graph} & fragmented benchmarks \\
Transformer / attention & GA inner-product attention, bilinear layers & meshes, proteins, HEP events & $E(3)$, Lorentz & GATr~\cite{brehmer2023geometric}, iP-/C-GATr~\cite{de2024euclidean}, L-GATr~\cite{spinner2024lorentz}, CFA~\cite{wagner2024generating} & memory and norm.\ cost \\
Generative / control & motors, versor actions, equivariant diffusion & poses, grasps, trajectories & $SE(3)$/$E(3)$ & CGA-PoseNet~\cite{pepe2024cgaposenet}, GAGrasp~\cite{zhong2025gagrasp}, hPGA-DP~\cite{sun2025hybrid} & narrow task scope \\
\bottomrule
\end{tabularx}
\end{table*}

\section{Applications}
\begin{figure*}[t]
\centering
\includegraphics[width=0.92\textwidth]{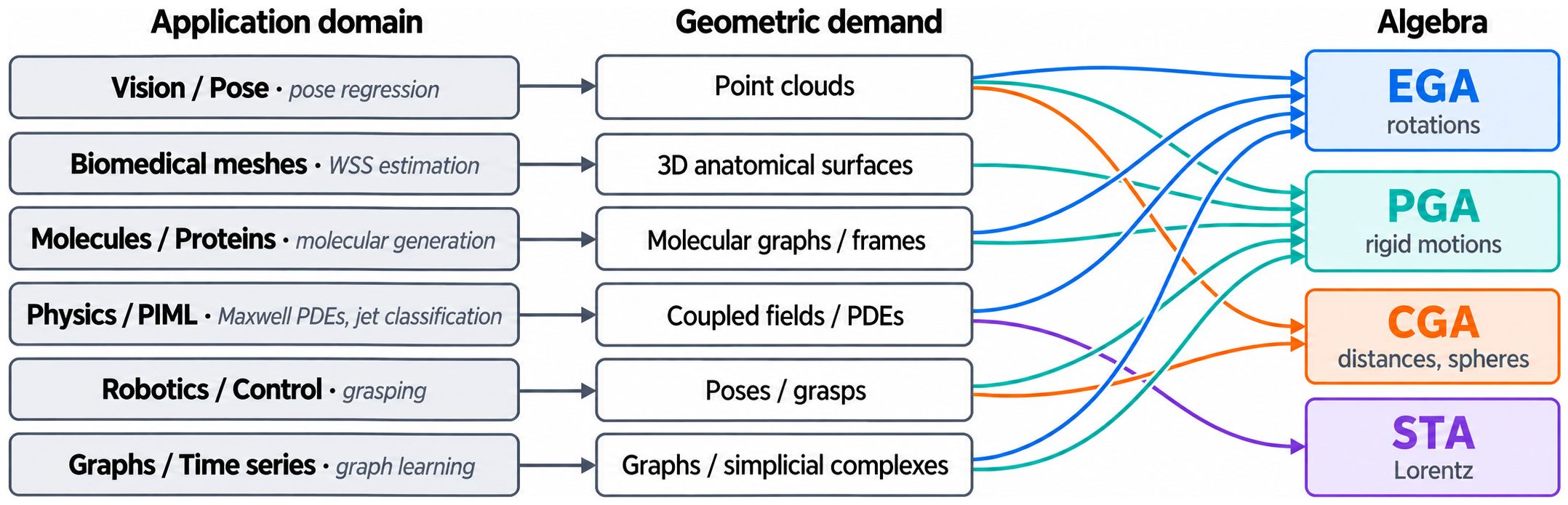}
\caption{Application-domain map for GA-native neural networks. The recurring pairings among domains, geometric demands, and algebra choices explain much of the current literature structure.}
\label{fig:application-map}
\vspace{-8pt}
\end{figure*}

The theoretical advantages of Geometric Algebra are increasingly translated into practical benefits across a diverse range of applications. This section highlights key application areas; Fig.~\ref{fig:application-map} and Table~\ref{tab:application_crosswalk} summarize how domains, symmetry demands, and algebra choices align across the literature.

\subsection{Computer Vision and Graphics}
Computer vision and graphics pipelines traditionally rely on disparate tools (homogeneous coordinates, quaternions, vector algebra), each with its own rules and limitations. GA unifies these, making it a natural substrate for geometrically-aware models~\cite{wareham2004applications,bayro2018geometric}. Many vision tasks reduce to regression on rotations, where GA rotors avoid gimbal lock and quaternion double cover. Pepe et al.~\cite{pepe2024cgaposenet,pepe2023cga} use CGA ($\mathbb{G}_{(4,0)}$) to represent $\mathrm{SE}(3)$ pose as a motor, an 8-parameter even-grade multivector encoding both rotation and translation, improving accuracy over Cartesian parametrizations. CGAPoseNet+GCAN~\cite{pepe2024cgaposenet} adds GCAN layers~\cite{ruhe2023geometric} for equivariance, further reducing errors with fewer parameters.

In 3D shape representation, GA-Planes~\cite{sivgin2024geometric} parameterizes volumetric features on multivector ``tensor planes,’’ becoming the first implicit neural volume model trainable via convex optimization, with competitive results on radiance field reconstruction, 3D segmentation, and video representation. Earlier work applied GA to indoor scene modeling~\cite{liu2018interactive} and UAV-based 3D reconstruction~\cite{khan2020clifford}. In graphics, Gunn~\cite{gunn2019course} shows that GA transformations yield more numerically stable shader programs than matrices, and Kolyvakis et al.~\cite{kolyvakis2025geometric} integrate CGA with language models for text-driven 3D scene editing.

GA is also effective for point cloud processing. CGA ($\mathbb{G}_{(4,1,0)}$) represents geometric primitives as single blades, enabling RANSAC-based detection of planes, spheres, and cylinders~\cite{kanatani2015understanding,du2017modeling,sveier2017object}. For learned features, Geometric Algebra Attention Networks~\cite{spellings2021geometric} generate rotationally invariant and covariant features via geometric products. CGENN~\cite{ruhe2023clifford} and GATr~\cite{brehmer2023geometric} treat points as tokens processed by fully equivariant layers, and recent work explores GA for generative tasks such as point cloud upsampling~\cite{xu2024point}.

Standard CNNs must learn rotation equivariance through augmentation. For vectorial or coupled data (optical flow, fluid fields, hyperspectral images), treating channels as independent scalars discards structural information. Clifford Convolutions~\cite{brandstetter2022clifford} and Clifford-Steerable CNNs~\cite{zhdanov2024clifford} generalize convolution to multivector fields, processing coupled quantities (e.g., scalar pressure and vector velocity) as unified objects via the geometric product. CS-CNNs achieve full $\mathrm{E}(p,q)$ equivariance by parameterizing steerable kernels with a CGENN-style MLP. For optical flow, GA-DDVM~\cite{pepe2025geometric} restricts diffusion to 2D vector fields with scaling and rotation operations, lowering KITTI and Sintel endpoint error relative to a matched DDVM baseline. CliffordNet~\cite{ji2026cliffordnet} restructures vision backbones using the geometric product as a unified interaction mechanism with $\mathcal{O}(N)$ complexity. In hyperspectral imaging, GA-CNN~\cite{li2022ga} represents spectral bands as multivector components for holistic cross-band learning, recently extended to anomaly detection~\cite{liu2025geometric}. RGA-CNNs~\cite{wang2021rga} similarly exploit geometric correlations for object detection and denoising.

\subsection{Computational Biology and Chemistry}
GA's 3D equivariance~\cite{suk2024geometric,suk2024lab,spellings2021geometric}, ability to encode higher-order geometry~\cite{de2024euclidean,wagner2024generating,pepe2024clifford,liu2025clifford}, and parameter efficiency~\cite{alblas2025geometric,liu2024multivector,alfarraj2022geometric} make it a natural fit for computational biology and chemistry. Multivector channels are also suited for large meshes, helping avoid oversquashing~\cite{suk2024lab}.

GA is well-suited for complex 3D biomedical meshes. The GATr architecture~\cite{brehmer2023geometric} has been applied to predict wall shear stress (WSS) on arterial walls. Following this, Suk et al.~\cite{suk2024lab} propose LaB-GATr for anatomical 3D meshes. It embeds meshes in PGA $\mathbb{G}_{3,0,1}$ and uses a GA tokenization module with message passing for efficient, global processing. This $E(3)$-equivariant model addresses canonical alignment issues across patients and was also applied to volume-based velocity field estimation and newborn postmenstrual age (PMA) estimation~\cite{suk2024lab}. LaB-GATr++~\cite{suk2024geometric} is an extension using cross-attention instead of message-passing pooling, achieving stronger accuracy and runtime on the same tasks. Separately, de Haan et al.~\cite{de2024euclidean} proposed several GATr variants (E-GATr, P-GATr, iP-GATr, and C-GATr) and tested them on the WSS prediction task~\cite{suk2024mesh}. They found all variants beat baselines, with the distance-aware C-GATr and iP-GATr performing best~\cite{de2024euclidean}. Based on LaB-GATr~\cite{suk2024geometric}, Alblas et al.~\cite{alblas2025geometric} develop a personalized growth prediction model for abdominal aortic aneurysms (AAAs), using local GA-embedded features to predict growth directly on the vascular surface. AB-GATr~\cite{rygiel2026when} is competitive with LaB-GATr at lower training cost and outperforms augmented non-equivariant hemodynamic baselines, though strong canonical alignment can favor non-equivariant models on ShapeNet-Car. Finally, Rao et al.~\cite{rao2024enhanced} use a GATr-like algorithm (SNGNN-BTGO) to extract features from MRI scans for brain tumor diagnosis, where GA's multidimensional representation helps distinguish between tumor types.

Another common application is molecular dynamics (MD)~\cite{chmiela2017machine}, where the goal is to predict molecule energy and trajectories. By preserving symmetries, GA-based models can simulate realistic thermodynamic behavior accurately and efficiently~\cite{spellings2021geometric, alfarraj2022geometric}. In early work, Spellings~\cite{spellings2021geometric} propose Geometric Algebra Attention Networks (GAANs), using a rotationally invariant and permutationally equivariant attention mechanism based on geometric products. For MD, this helps learn a conservative force field. Extending CGENNs~\cite{ruhe2023clifford}, Liu et al.~\cite{liu2024clifford} introduce CSMPNs, which learn to embed higher-order simplices via geometric products and equivariant simplicial message passing~\cite{huang2024higher}. Similarly, Tran et al.~\cite{tran2024clifford} propose CG-EGNNs, which combine a GA embedding layer with Clifford graph convolution layers. This equivariant model reports low MSE on the MD task, especially for complex cyclic molecules, highlighting GA's value in capturing high-order information~\cite{tran2024clifford}. More recently, GA-GNN~\cite{petersen2026design} embeds all node attributes into a single multivector representation and relies entirely on geometric product layers as the core update mechanism. It exhibits high accuracy on targets with clear geometric interpretations, such as the dipole moment and electronic spatial extent. CliffordIP~\cite{polat2026cliffordip} builds $\mathrm{O}(3)$-equivariant $\mathbb{G}_{3,0,0}$ interatomic potentials for heterogeneous catalysis without spherical harmonics, reducing energy MAE by 20--24\% over the next-best baseline on adsorbate subsets.

Beyond MD, Alfarraj et al.~\cite{alfarraj2022geometric} use GA to generate molecular surfaces. Their method uses the \textit{Clifford-Fourier transform}~\cite{hitzer2002multivector}, implemented via a \textit{Fast Clifford-Fourier transform (FCFT)}, to produce surfaces free of geometric singularities. For 3D molecular generation, Liu et al.~\cite{liu2025clifford} propose Clifford Diffusion Models (CDMs). CDMs use Clifford-EGNNs~\cite{liu2024multivector} as latent encoders and perform diffusion and denoising entirely in GA space, operating on all multivector subspaces to capture rich geometric information and improve sample stability.

A prominent application is protein design and structure, where GA provides an efficient framework to describe amino acid positions and orientations using higher-order multivector subspaces~\cite{pepe2024clifford}. Leveraging CGENN layers~\cite{ruhe2023clifford}, Pepe et al.~\cite{pepe2024clifford} apply Euclidean GA to protein structure prediction (PSP), using MVL and FCGP layers to project protein graphs. This avoids "flattening" geometry. The same authors also used CGA for PSP~\cite{10.1007/978-3-031-55985-3_6}, modeling the backbone in CGA and applying their architecture to predict cost maps, which measure relative amino acid orientations. Hockey et al.~\cite{hockey2024simplifying} introduce CGA Transformer, a GATr variant, for PSP, specifically to predict amino acid inter-residue distances without directly predicting coordinates. To avoid expensive geometric product layers~\cite{ruhe2023clifford}, Liu et al.~\cite{liu2024multivector} propose cheaper, MLP-based models: Clifford-EGNN (an extension of EGNN~\cite{satorras2021n}), MVN-GNN (using multivector neurons), and MVP-GNN (using multivector perceptrons). These models were applied to protein structure denoising. Spellings~\cite{spellings2021geometric} also applies GAANs to protein coarse-graining backmapping. This task reverses coarse-graining to recover original coordinates~\cite{marrink2013perspective}. GAANs use permutation-covariant and rotation-covariant layers to predict heavy atom coordinates from coarse-grained centers of mass.

A significant application is generative protein backbone design. Wagner et al.~\cite{wagner2024generating} introduce Geometric Algebra Flow Matching (GAFL) for protein backbone design. It replaces the invariant point attention (IPA) from AlphaFold2~\cite{jumper2021highly} with Clifford Frame Attention (CFA). CFA uses PGA to represent the backbone, secondary structures, and their relations, replacing point-valued attention with PGA multivectors. It also uses geometric bilinear layers~\cite{brehmer2023geometric} for message passing. GAFL is tested on a backbone generation task, showing good designability, diversity, and novelty. Additionally, Alfarraj et al.~\cite{alfarraj2023geometric} propose a GA-based method for protein-protein docking~\cite{vakser2014protein}, finding that rotors and the geometric product can express rotations more efficiently than traditional methods. Beyond proteins, RBFN~\cite{wang2025geometric} introduces a multivector parameterization for RNA, which conditions a Bayesian Flow Network with a novel time-step sampling distribution. This geometric enhancement allows RBFN to effectively align the sequence distribution with the underlying RNA backbone structure, achieving superior sequence recovery rates over both standard deep learning baselines (e.g., gRNAde~\cite{joshi2024grnade}) and physics-based tools like Rosetta~\cite{das2008macromolecular}.

\subsection{Physics-Informed Machine Learning}
Physics-Informed Machine Learning (PIML)~\cite{karniadakis2021physics,hao2022physics} seeks models consistent with known physical laws. GA-based PIML models embody a strong form of this prior: rather than treating coupled fields (e.g., scalar pressure $p(x)$ and vector velocity $u(x)$, or electric $E$ and magnetic $B$ fields) as separate channels, GA represents the entire physical state as a unified multivector field~\cite{brandstetter2022clifford}. GA-based layers then process this object intrinsically, providing a deep architectural inductive bias beyond surface-level loss regularization.

In the field of computational fluid dynamics (CFD), initial work focuses on developing GA-based layers as direct, geometrically-aware replacements for standard NN components. Brandstetter et al.~\cite{brandstetter2022clifford} introduce Clifford Convolutions and Clifford Fourier Transforms to replace standard layers in neural PDE surrogates for 2D Navier-Stokes equations and weather modeling. Similarly, GCAN~\cite{ruhe2023geometric} demonstrates strong performance on modeling shallow water equations and incompressible Navier-Stokes. This field has rapidly evolved from GA-based layers to full-fledged GA-based neural operators. The Fengbo~\cite{pepefengbo} model, for instance, is a Clifford Neural Operator pipeline designed for 3D CFD. It conceptualizes the problem as an interpretable mapping from the geometry of the domain (represented as a multivector) to the physics of the solution (the resulting multivector-valued pressure and velocity field). It is built on Clifford-FNO layers and achieves competitive accuracy on 3D CFD benchmarks with significantly fewer parameters than competing graph-based methods.

The domain of electromagnetism and relativistic physics also provides compelling evidence for the GA-NN thesis, as it demonstrates that the choice of algebra is a critical, non-trivial design decision. STAResNet~\cite{pepe2024staresnet} directly tackles the problem of solving Maxwell's partial differential equations. Pepe et al.~\cite{pepe2024staresnet} compare a standard Clifford ResNet using Euclidean GA ($\mathbb{G}_{3,0,0}$) against their STAResNet, which operates natively in STA. The Euclidean model must learn from the "split" 3D vector and scalar time representation, whereas STAResNet operates directly on the unified STA multivector $F = E + IB$ and achieves a lower error with significantly fewer parameters. This principle extends directly to high-energy physics, where Lorentz covariance is the fundamental symmetry. Spinner et al.~\cite{spinner2024lorentz} propose L-GATr that operates entirely in Spacetime Algebra ($\mathbb{G}_{1,3}$) to achieve full Lorentz equivariance. It has been successfully applied to diverse Large Hadron Collider (LHC) tasks, including amplitude regression and jet classification~\cite{brehmer2024lorentz}. It was also used to construct the first Lorentz-equivariant generative model for particle physics.

\subsection{Robotics}
Classical robotics uses CGA for closed-form inverse kinematics of manipulators with spherical wrists~\cite{zaplana2022closed}, including 6-DOF and 7-DOF redundant arms~\cite{fu2013solution}, where forward kinematics is expressed as a chain of motors~\cite{zhu2025conformal}. This framework has been extended to optimal control~\cite{loewGeometricAlgebraOptimal2023}, but the limitations of analytic solvers~\cite{carbajal2015modeling} motivate learning-based alternatives. Early work by Bayro-Corrochano et al.~\cite{bayro2018geometric} developed quaternion spiking neural networks for robot motion interpolation, but these were relatively small-scale. Recent GA-NNs leverage equivariance for data-efficient deep robot learning. GAGrasp~\cite{zhong2025gagrasp} uses PGA ($\mathbb{G}_{3,0,1}$) to represent $\mathrm{SE}(3)$ transformations in a diffusion model for dexterous grasp generation, achieving robust generalization to novel poses while significantly outperforming non-equivariant baselines. GATr~\cite{brehmer2023geometric} serves as the backbone for $E(3)$-invariant motion planning, and hPGA-DP~\cite{sun2025hybrid} integrates PGA-GATr into visuomotor diffusion policies for faster convergence and improved task performance. For self-driving, DriveGATr~\cite{xu2026efficient} encodes scene elements in 2D PGA ($\mathbb{G}_{2,0,1}$) for $\mathrm{SE}(2)$-equivariant agent modeling, replacing costly pairwise relative positional encodings with standard multivector attention.

\subsection{Other Notable Applications}
GA-based networks extend beyond geometry-heavy domains wherever data has multi-dimensional or relational structure. In graph representation learning, GGAN~\cite{zhong2025graph} generalizes message passing to a GA-defined feature space, encoding node attributes as multivectors and reporting improved expressiveness with reduced model complexity. GA embeddings have also been applied to cross-domain few-shot graph classification~\cite{liu2022geometric}, where the geometric richness helps reduce feature distortion at depth.

In temporal and signal processing, GA-LSTNet~\cite{li2022geometric} represents each time-step as a multivector encapsulating all sensor channels, modeling joint dynamics more effectively than channel-wise models. Parametrized Hypercomplex NNs~\cite{grassucci2022phnns} provide a unified framework that automatically infers the optimal algebra (complex, quaternion, or higher-dimensional Clifford) for a given dataset, operating with fewer parameters. Wi-GATr~\cite{hehn2025differentiable} applies GATr to wireless signal propagation, predicting channel observations from scene meshes and antenna poses, with lower real-world error than hybrid methods and a calibrated ray tracer.

Emerging interdisciplinary applications include CGA-based interfaces with language models for spatial instruction execution~\cite{kolyvakis2025geometric}, GA multivectors as word embeddings~\cite{mani2023representing}, randomized GA methods for LLM fine-tuning~\cite{wang2024randomized}, and GA networks for IoT anomaly detection on microcontrollers~\cite{garcia2022embedded}.

\begin{table*}[t]
\caption{Representative applications of GA-native neural networks, organized by domain. Cited works are representative.}
\label{tab:application_crosswalk}
\centering
\footnotesize
\setlength{\tabcolsep}{4pt}
\renewcommand{\arraystretch}{1.15}
\begin{tabularx}{\textwidth}{@{}l>{\raggedright\arraybackslash\hsize=.9\hsize}X>{\raggedright\arraybackslash}p{1.5cm}>{\raggedright\arraybackslash\hsize=1.0\hsize}X>{\raggedright\arraybackslash\hsize=1.1\hsize}X@{}}
\toprule
Domain & Geometry / symmetry & Algebra(s) & Canonical tasks & Representative works \\
\midrule
Vision / pose & point clouds, multi-view; rotation / rigid motion & EGA, PGA, CGA & camera pose regression, implicit volumes, hyperspectral classification & CGA-PoseNet~\cite{pepe2024cgaposenet}, GA-Planes~\cite{sivgin2024geometric}, GA-CNN~\cite{li2022ga} \\
Biomedical meshes & 3D anatomical surfaces; $E(3)$ equivariance & PGA & wall shear stress estimation, aortic aneurysm growth, velocity field prediction & GATr~\cite{brehmer2023geometric}, LaB-GATr(++)~\cite{suk2024lab,suk2024geometric}, iP-/C-GATr~\cite{de2024euclidean}, AAA model~\cite{alblas2025geometric}, AB-GATr~\cite{rygiel2026when} \\
Molecules / proteins & molecular graphs, residue frames; $E(3)$, higher-order geometry & PGA, EGA & molecular dynamics (MD17), protein denoising, backbone generation & CSMPN~\cite{liu2024clifford}, GAFL~\cite{wagner2024generating}, CDM~\cite{liu2025clifford}, GA-GNN~\cite{petersen2026design}, CliffordIP~\cite{polat2026cliffordip} \\
Physics / PIML & coupled fields, PDE states; field / Lorentz symmetry & EGA, STA & Navier--Stokes surrogates, Maxwell PDEs, LHC tasks & Clifford layers~\cite{brandstetter2022clifford}, STAResNet~\cite{pepe2024staresnet}, L-GATr~\cite{spinner2024lorentz} \\
Robotics / control & poses, grasps; $SE(3)$ & PGA, CGA & grasp generation, motion planning, visuomotor policy & GAGrasp~\cite{zhong2025gagrasp}, hPGA-DP~\cite{sun2025hybrid} \\
Graphs / time series & graphs, simplicial complexes, multivariate series & EGA, PGA & graph learning, few-shot transfer, forecasting & GGAN~\cite{zhong2025graph}, CG-EGNN~\cite{tran2024clifford}, GA-LSTNet~\cite{li2022geometric} \\
\bottomrule
\end{tabularx}
\vspace{-5pt}
\end{table*}

\section{Empirical Studies}
The empirical literature on GA-based neural networks is highly heterogeneous. Architectures differ in algebra, symmetry target, data modality, output type, and metric, so a single leaderboard would mix incomparable settings. This section does not propose such a ranking. Instead, we summarize evidence only within matched experimental settings and distinguish recurring cross-paper trends from isolated positive reports. Because baselines, training budgets, and implementation quality differ substantially across papers, cross-paper comparisons should be interpreted cautiously; empirical gains reported for GA-native models should be weighed relative not only to non-geometric baselines, but also to strong non-GA symmetry-aware alternatives when such comparisons are available. Table~\ref{tab:empirical_matrix} summarizes this evidence structure.

\begin{table*}[t]
\caption{Empirical evidence map for GA-native neural networks, organized by matched benchmark setting. Each row is a within-task comparison unit; cross-row comparisons are intentionally avoided. The \emph{Evidence} column grades each setting as \textbf{Moderate} (replicated gains on established benchmarks, but confounds or limited independent replication remain) or \textbf{Preliminary} (too few matched studies or too recent for confident conclusions). No setting currently reaches ``Strong'' by strict archival-survey standards.}
\label{tab:empirical_matrix}
\centering
\footnotesize
\setlength{\tabcolsep}{4pt}
\renewcommand{\arraystretch}{1.15}
\begin{tabularx}{\textwidth}{@{}>{\raggedright\arraybackslash}p{3.1cm}>{\raggedright\arraybackslash\hsize=1.1\hsize}X>{\raggedright\arraybackslash}p{2.7cm}>{\raggedright\arraybackslash\hsize=.9\hsize}X>{\raggedright\arraybackslash}l>{\raggedright\arraybackslash}p{2.1cm}@{}}
\toprule
Matched setting & GA works & Comparator class & Strongest supported claim & Evidence & Scope limit \\
\midrule
Synthetic $n$-body dynamics & GATr~\cite{brehmer2023geometric}, iP-/C-GATr~\cite{de2024euclidean}, GCAN~\cite{ruhe2023geometric} & SEGNN, $SE(3)$-Transf. & sample efficiency; robustness to translation shift & Moderate & synthetic only \\
Arterial WSS / biomed.\ meshes & GATr~\cite{brehmer2023geometric}, LaB-GATr(++)~\cite{suk2024lab,suk2024geometric}, iP-/C-GATr~\cite{de2024euclidean}, AB-GATr~\cite{rygiel2026when} & mesh GDL, algebra ablations & accuracy on non-canon.\ meshes; sample efficiency & Moderate & biomed.\ mesh only \\
Navier--Stokes / weather / Maxwell & Clifford layers~\cite{brandstetter2022clifford}, GCAN~\cite{ruhe2023geometric}, STAResNet~\cite{pepe2024staresnet}, C-CSCNN~\cite{szarvas2026conditional} & CNN, FNO, operator baselines & param.\ efficiency; matched-algebra advantage & Moderate & PDE-specific \\
HEP regression / classif.\ / generation & L-GATr~\cite{spinner2024lorentz,brehmer2024lorentz,favaro2026virtues} & Lorentz-equiv.\ and standard HEP baselines & competitive or superior accuracy; Lorentz faithfulness & Moderate & HEP only \\
MD17 / protein denoising / higher-order & CSMPN~\cite{liu2024clifford}, CG-EGNN~\cite{tran2024clifford}, MVN~\cite{liu2024multivector} & EGNN, higher-order GNNs & higher-order geometry benefit; negative result on benzene & Moderate & benchmark-local \\
Protein backbone generation & GAFL / CFA~\cite{wagner2024generating} & structure-aware protein baselines & designability, diversity, novelty & Prelim. & not comparable to prediction \\
Camera pose regression & CGA-PoseNet~\cite{pepe2024cgaposenet} & pose-regression baselines & accuracy, parameter efficiency & Prelim. & pose studies only \\
Grasp / visuomotor control & GAGrasp~\cite{zhong2025gagrasp}, hPGA-DP~\cite{sun2025hybrid} & diffusion / policy baselines & data efficiency, $SE(3)$ generalization & Prelim. & robotics; task-local \\
\bottomrule
\end{tabularx}
\vspace{-8pt}
\end{table*}

\subsection{What transfers across the literature}
The strongest recurring evidence concerns symmetry handling. Architectures such as CGENNs~\cite{ruhe2023clifford}, GATr~\cite{brehmer2023geometric}, L-GATr~\cite{spinner2024lorentz}, and CSMPNs~\cite{liu2024clifford} are equivariant by construction under the relevant Clifford-group, Euclidean, or Lorentz action. Empirically, these guarantees matter most under transformation shift. In GATr~\cite{brehmer2023geometric}, performance is retained on translated $n$-body systems, whereas non-equivariant baselines degrade markedly. On arterial wall-shear-stress estimation, GATr~\cite{brehmer2023geometric} is strongest on non-canonicalized meshes in the benchmark of Suk et al.~\cite{suk2024mesh}, where the model must handle arbitrary orientation directly.

Data efficiency is the second consistent theme. GATr~\cite{brehmer2023geometric} is more sample-efficient than SEGNN and the $\mathrm{SE}(3)$-Transformer on synthetic $n$-body data, while iP-GATr and C-GATr~\cite{de2024euclidean} improve further on the same family of tasks. GCAN~\cite{ruhe2023geometric} reports similar low-data gains on dynamical systems. In robotics, GAGrasp~\cite{zhong2025gagrasp} and hPGA-DP~\cite{sun2025hybrid} report faster convergence or stronger low-data behavior, though only within their own benchmarks.

A third pattern is geometric expressivity. Clifford neural layers~\cite{brandstetter2022clifford} improve PDE surrogates by processing coupled scalar and vector fields as multivectors rather than as unrelated channels. On WSS estimation, the strongest GATr variants are precisely those with richer geometry-aware interactions, notably distance-aware attention in iP-GATr and C-GATr~\cite{de2024euclidean}. In higher-order graph learning, CSMPN~\cite{liu2024clifford} benefits from building simplex features algebraically via geometric products.

Interpretability remains the least established claim. CGENN-based protein models~\cite{pepe2024clifford} and CGA-PoseNet variants~\cite{pepe2024cgaposenet} argue that motors, residue frames, or grade-structured channels are easier to inspect than unconstrained Euclidean features, but the field still lacks controlled interpretability benchmarks. At present, interpretability is better framed as a plausible representational advantage than as settled empirical evidence.

\subsection{Comparative evidence by application family}
\paragraph{Vision, graphics, and mesh-based geometry.}
This is the clearest shared-task setting. The arterial WSS benchmark of Suk et al.~\cite{suk2024mesh} appears in mesh neural networks, the original GATr work~\cite{brehmer2023geometric}, the algebra-selection study of de Haan et al.~\cite{de2024euclidean}, and the LaB-GATr line~\cite{suk2024lab,suk2024geometric}. Across these studies, the strongest support is for predictive accuracy on randomly oriented meshes, robustness to non-canonical orientation, and scalability to larger biomedical meshes.

\paragraph{Biology, chemistry, proteins, and molecules.}
Evidence here is broader but less standardized. CSMPN~\cite{liu2024clifford} improves over simplicial and equivariant graph baselines on convex-hull and trajectory tasks, but also includes a useful negative result: for rigid planar benzene, extra geometric capacity does not help uniformly. Multivector Neurons~\cite{liu2024multivector} and CG-EGNN~\cite{tran2024clifford} report gains on $n$-body, protein denoising, CMU motion, and MD17, though this evidence is still mostly preprint-era. In protein structure prediction, replacing linear layers with CGENN layers~\cite{pepe2024clifford} improves the tested pipeline by up to $2.1\%$ while using fewer parameters. GAFL~\cite{wagner2024generating} reports strong generative metrics for backbone design.

\paragraph{Physics, PDEs, operators, and high-energy physics.}
In physics-informed settings, the key question is often whether the \emph{matched} algebra helps. Clifford neural layers~\cite{brandstetter2022clifford} improve generalization on Navier--Stokes, weather, and Maxwell benchmarks at similar parameter counts. GCAN~\cite{ruhe2023geometric} reports analogous gains on rigid-body and fluid-dynamics problems. STAResNet~\cite{pepe2024staresnet} is especially informative: under a controlled comparison, the STA variant reaches up to $2.6\times$ lower MSE with $6\times$ fewer parameters than a Euclidean-GA ResNet on Maxwell prediction. In high-energy physics, L-GATr~\cite{spinner2024lorentz,brehmer2024lorentz} is competitive with or stronger than strong domain baselines on amplitude regression, jet classification, and Lorentz-equivariant generation. Large-scale jet-tagging studies find that Lorentz-equivariant transformers outperform standard ones when geometric features matter~\cite{favaro2026virtues}, and on precision unfolding and anomaly-detection tasks L-GATr performs comparably to the pretrained foundation model OmniLearn~\cite{bresopla2026encoding}.

\paragraph{Robotics, control, and manipulation.}
Robotics evidence is positive but fragmented. GAGrasp~\cite{zhong2025gagrasp} reports better pose generalization and stronger efficiency for dexterous grasp generation under $\mathrm{SE}(3)$ transformations. hPGA-DP~\cite{sun2025hybrid} reports faster convergence and better task performance than standard diffusion policies and architectures that rely only on P-GATr. These results support data efficiency and robustness, but not yet a fair cross-paper ranking.

\paragraph{Other graph and signal domains.}
A smaller set of papers extends GA-based learning beyond geometry-heavy domains. GGAN~\cite{zhong2025graph} reports strong graph-representation-learning results with reduced model complexity, and GA-LSTNet~\cite{li2022geometric} argues that multivector recurrent states improve multivariate time-series modeling by preserving inter-channel structure. These results are promising, but the evidence base remains comparatively thin.

\subsection{Computational cost and scalability}
A recurring cost of GA is dimensional overhead: dense multivectors scale with the number of basis blades, which grows exponentially with algebra dimension. In practice, this means wider hidden states, denser bilinear products, costlier normalization, and a larger implementation burden. The algebra-selection study of de Haan et al.~\cite{de2024euclidean} makes the trade-off explicit: E-GATr is cheapest but weaker, simple P-GATr is under-expressive, and C-GATr is strong but harder to stabilize.

GA models are not uniformly unscalable. GATr~\cite{brehmer2023geometric} shows that a GA model can inherit Transformer-like scaling when its core interaction is efficient attention rather than dense message passing. LaB-GATr~\cite{suk2024lab,suk2024geometric} extends this to large biomedical meshes by compressing token sets while preserving Euclidean equivariance. By contrast, dense GA operations remain expensive in some PDE~\cite{ruhe2023geometric} and graph~\cite{liu2024clifford} settings.

\subsection{Failure modes and limitations}
We organize observed limitations into three categories: evidence quality, model and training failure modes, and practical deployment constraints.

\paragraph{Evidence-quality limitations.}
Benchmark fragmentation remains severe: many papers evaluate on one or two specialized tasks with metrics and preprocessing that differ substantially across studies, making broad claims easy to overstate. Reproducibility is uneven, with several influential results existing only as preprints or workshop papers. Apparent gains can narrow or vanish under compute-, memory-, or latency-matched comparisons, since lower parameter counts do not guarantee faster wall-clock throughput.

\paragraph{Model and training failure modes.}
Performance is highly sensitive to algebra choice: if the selected algebra or metric does not match the task geometry, GA introduces computational overhead without a commensurate inductive-bias benefit, and mature non-GA equivariant baselines may remain preferable; the benzene result in CSMPN~\cite{liu2024clifford} is a concrete example. Expressivity can also be structurally limited: vector--vector geometric products in $\mathbb{G}_{3,0,0}$ lack the rank-2 ($L=2$) component, which limits force-direction accuracy in tested Clifford interatomic potentials~\cite{polat2026geometric}. Optimization difficulties extend beyond isolated normalization issues. In practice, multivector activations can collapse toward a subset of grades (grade collapse), constrained geometric objects such as rotors or versors may drift from their intended manifolds unless explicitly re-normalized, and initialization sensitivity can be acute. These problems are exacerbated in indefinite algebras such as CGA and STA, where the C-GATr study~\cite{de2024euclidean} reports that non-positive norms and cancellation effects make standard deep-learning heuristics less reliable and often necessitate full 32-bit precision. Equivariance can also leak in practice through preprocessing choices, discretization, neighborhood construction, or coordinate conventions that inadvertently break the intended algebraic structure.

\paragraph{Practical deployment constraints.}
Geometric products and grade-coupled operations increase memory traffic and interact poorly with mixed-precision pipelines and current GPU kernel libraries, limiting throughput even when parameter counts are competitive. Nevertheless, slim and quantized L-GATr jet taggers are estimated to cut inference energy by an order of magnitude at a moderate performance loss~\cite{petitjean2025economical}. For homogeneous point-feature tasks or dense 2D image/video pipelines, GA should currently be viewed as a specialized option rather than a default choice. Evidence on robustness under realistic perception perturbations (sensor noise, calibration error, partial observability, and distribution shift) remains limited. The strongest results arise when the task genuinely requires joint reasoning over heterogeneous geometric entities, frames, or coupled physical quantities.

\subsection{Summary of evidence status}
Across the settings surveyed above, the current evidence base can be summarized as follows:
\begin{itemize}[leftmargin=*,itemsep=1pt]
\item \textbf{Established:} improved inductive alignment for tasks with explicit geometric symmetries, transformations, or multi-entity geometric relations.
\item \textbf{Plausible but not yet general:} gains in data efficiency, robustness under distribution shift, and compactness through structured parameterization.
\item \textbf{Currently unproven:} broad superiority on dense perception, large-scale vision backbones, or standardized foundation-model-scale settings.
\end{itemize}

\subsection{Recommended benchmarking practice}
Future GA-NN papers should adopt the following protocol to make local comparisons cumulative without forcing the field into a misleading global leaderboard.
\begin{enumerate}[leftmargin=*,itemsep=1pt]
    \item \textbf{State the algebraic hypothesis.} Specify the algebra, target symmetry group, and whether equivariance is exact or approximate.
    \item \textbf{Use matched baselines.} Compare against both non-equivariant and strong equivariant baselines under matched parameter counts and training budgets.
    \item \textbf{Include symmetry stress tests.} Report transformed-input evaluations and state whether canonicalization was required.
    \item \textbf{Report low-data behavior.} Include learning curves or sub-sampling experiments when sample efficiency is claimed.
    \item \textbf{Report cost alongside accuracy.} At minimum: parameter count, training precision, wall-clock time, and peak memory.
    \item \textbf{Ablate the geometry.} Remove multivector channels, replace geometric products with linear mixing, or swap to a mismatched algebra.
    \item \textbf{Release reproducibility materials.} Code, splits, preprocessing, and evaluation scripts are essential for custom pipelines.
\end{enumerate}

\section{Software Libraries and Implementation}
GA neural networks rely on software libraries to represent multivectors, implement the geometric product and related Clifford operations, and integrate these objects into deep learning stacks. Recent libraries span Python, C++, Java, JavaScript, PyTorch, TensorFlow, and JAX. Table~\ref{tab:software_ecosystem} summarizes the main ecosystem relevant to GA-native learning; the text below highlights representative tools and current bottlenecks.

\begin{table*}[t]
\caption{Software ecosystem relevant to GA-native neural networks (as of 2026).
``DL'' marks libraries with native deep-learning tensor support;
the remainder are usable for preprocessing or non-DL computation.}
\label{tab:software_ecosystem}
\centering
\footnotesize
\setlength{\tabcolsep}{4pt}
\renewcommand{\arraystretch}{1.15}
\begin{tabular}{@{}>{\raggedright\arraybackslash}p{2.2cm}
                >{\raggedright\arraybackslash}p{2.45cm}
                c
                >{\raggedright\arraybackslash}p{3.4cm}
                >{\raggedright\arraybackslash}p{5.0cm}@{}}
\toprule
Library & Lang.\ / Framework & DL & Algebra scope & Primary use case \\
\midrule
\multicolumn{5}{@{}l}{\textit{ML-facing libraries}} \\
\midrule
torch\_ga~\cite{torch_ga}       & Python / PyTorch     & $\surd$ & tensor multivectors              & rapid GA layer prototyping \\
tfga~\cite{python_tfga}         & Python / TensorFlow  & $\surd$ & tensor multivectors              & Keras-style GA experiments \\
numga~\cite{numga}              & Python / JAX         & $\surd$ & symbolic + numerical, ${\leq}6$D & operator-fused algebra kernels \\
lgatr~\cite{lgatr}              & Python / PyTorch     & $\surd$ & STA $\mathbb{G}_{1,3,0}$ (L-GATr) & Lorentz-equivariant transformers for HEP \\
\midrule
\multicolumn{5}{@{}l}{\textit{General-purpose scripting tooling}} \\
\midrule
clifford~\cite{python_clifford} & Python               &         & broad common algebras            & preprocessing, general GA computation \\
Kingdon~\cite{roelfs2025willingkingdoncliffordalgebra} & Python & & symbolic + numerical             & mixed symbolic / numerical prototyping \\
ganja.js~\cite{ganja.js}        & JavaScript           &         & code-generated, broad            & pedagogy, visualization, code export \\
\midrule
\multicolumn{5}{@{}l}{\textit{Compiled / systems libraries}} \\
\midrule
Versor~\cite{colapinto2011versor} & C++               &         & EGA, PGA, CGA, STA               & performance-sensitive geometry engines \\
GATL~\cite{fernandes-SEMA-13-2021} & C++             &         & arbitrary algebras               & compile-time-optimized kernels \\
Klein~\cite{Klein}              & C++                  &         & 3D PGA only                      & SIMD-optimized rigid-body PGA \\
Gaalop~\cite{Gaalop}            & Java / codegen       &         & compiled GA expressions          & high-performance generated kernels \\
Garamon~\cite{breuils_garamon_2019} & C++ generator   &         & user-defined algebras            & custom or uncommon algebras \\
TbGAL~\cite{sousa2020tbgal}     & C++ tensor-based     &         & blade / versor repr.             & high-dimensional performance studies \\
\bottomrule
\end{tabular}
\vspace{-8pt}
\end{table*}

\subsection{ML Libraries: PyTorch, TensorFlow, and JAX}
The primary ML-facing libraries are \texttt{torch\_ga}~\cite{torch_ga} (PyTorch) and \texttt{tfga}~\cite{python_tfga} (TensorFlow), which both represent multivectors as framework tensors and provide Keras-style layers such as \texttt{GeometricSandwichProductDense} (analogous to Keras's \texttt{Dense}). Both libraries feature a \texttt{GeometricAlgebra} class supporting low-level tensor operations and higher-level \texttt{Multivector} abstractions with operator overrides. \texttt{numga}~\cite{numga} targets JAX (with NumPy/PyTorch fallbacks) and symbolically fuses GA operations into optimized linear operators at compile time; for example, it can deduce that the product of two quaternions remains a quaternion. However, numga is limited to algebras of dimension $\leq 6$ due to computational overhead.

\subsection{General Purpose Libraries}
Among scripting tools, ganja.js~\cite{ganja.js} is a lightweight JavaScript code generator with visualization and cross-language export; clifford~\cite{python_clifford} is a mature Python numerical toolkit; and Kingdon~\cite{roelfs2025willingkingdoncliffordalgebra} adds symbolic computation accepting numpy, PyTorch, and sympy. Gaalop~\cite{Gaalop,hildenbrand2010gaalop} compiles GA expressions into optimized C++, OpenCL, or CUDA code.

In C++, Versor~\cite{colapinto2011versor} and GATL~\cite{fernandes-SEMA-13-2021} use template meta-programming for compile-time optimization; Klein~\cite{Klein} provides SIMD-optimized 3D PGA; Garamon~\cite{breuils_garamon_2019} generates code for user-defined algebras; and TbGAL~\cite{sousa2020tbgal} uses tensor-based blade representations for higher-dimensional performance. Benchmarking studies~\cite{fernandes2021exploring,sousa2020tbgal} show GATL and Versor generally perform best, depending on operation mix and algebra dimension.
    
\subsection{Challenges and Future Needs for Software Support}
The $2^n$ scaling of basis blades limits most libraries to moderate dimensions: Gaalop~\cite{Gaalop, hildenbrand2010gaalop}, Garamon~\cite{breuils_garamon_2019}, GluCat~\cite{GluCat}, and Versor~\cite{colapinto2011versor} support at most $16$-, $20$-, $16$-, and $7$-dimensional spaces, respectively. Highly optimized libraries like Klein~\cite{Klein} sacrifice generality. Alternatives such as TbGAL's tensor-based approach~\cite{sousa2020tbgal} and custom GPU kernels~\cite{ruhe2023clifford} are promising for higher-dimensional algebras.

ML-facing libraries such as tfga~\cite{python_tfga} (unmaintained since 2023) and torch\_ga~\cite{torch_ga} are immature: APIs remain rough, and higher-dimensional convolutions are not yet supported. More broadly, GA's steep learning curve, inconsistent cross-library APIs, sparse documentation, and limited benchmarking~\cite{fernandes2021exploring,sousa2020tbgal} remain barriers to wider adoption.

\section{Future Directions and Conclusion}
GA-native neural networks have progressed from conceptual proposals to practical architectures with promising results in several geometry-centric domains, including high-energy physics, robotics, and computational biology. Empirical gains are most consistent when the task requires joint reasoning over heterogeneous geometric entities and explicit symmetries; for homogeneous point-feature or dense 2D settings, mature non-GA alternatives remain competitive. Several frontiers remain.

\paragraph{Algebra matching and numerical stability.}
Systematically mapping problem symmetries to the optimal algebra remains open. Indefinite algebras (CGA, STA) can break standard components like LayerNorm~\cite{de2024euclidean}; numerically robust layers for these spaces are critical.

\paragraph{Scalability and architectural depth.}
Dense multivector representations scale as $2^n$, creating a tension between expressivity and feasibility that calls for sparse representations and custom GPU kernels~\cite{ruhe2023clifford}. Current non-linearities are typically scalar-gated; fully algebra-aware activations and richer GA attention mechanisms~\cite{de2024euclidean} remain underexplored.

\paragraph{Interpretability and foundation-model integration.}
GA representations are argued to be more inspectable, but controlled benchmarks are lacking. GA could serve as a geometric interface for embodied AI and language models~\cite{kolyvakis2025geometric}, mapping symbolic instructions into continuous transformations.

\paragraph{Standardized evaluation.}
Benchmarking fragmentation (Section~VI) is arguably the largest barrier to cumulative progress. A shared evaluation suite spanning mesh regression, molecular dynamics, PDE surrogates, pose estimation, and robotic control, with standardized metrics and cost reporting, would significantly accelerate the field.

Multiple GA-native architecture families now deliver reproducible gains on real-world tasks, supported by a growing software ecosystem, but larger-scale and more reproducible evidence is still needed before GA can be considered a standard tool. The central remaining challenges (algebra selection, numerical stability in indefinite metrics, scalable implementations, and reproducible benchmarking) are increasingly concrete; addressing them will determine whether GA becomes a routine component of the geometric deep learning toolkit. For the pattern analysis community, the most promising near-term opportunities lie in 3D vision, embodied perception, multi-sensor fusion, and manipulation, where heterogeneous geometric entities and explicit symmetries are central. We hope this survey provides both a practical reference for researchers entering the field and a concrete roadmap for the community.

\bibliographystyle{IEEEtran}
\bibliography{main}

\end{document}